\documentclass[aps,prd,onecolumn,superscriptaddress,amsmath,amssymb,longbibliography]{revtex4-2}
\usepackage[utf8]{inputenc}
\usepackage[T1]{fontenc}
\usepackage{graphicx}
\usepackage{booktabs}
\usepackage[colorlinks=true,citecolor=blue,linkcolor=blue,urlcolor=blue]{hyperref}
\usepackage{xcolor}
\usepackage{caption}
\usepackage{subcaption}
\usepackage{siunitx}
\usepackage{multirow}

\newcommand{\eff}{{\rm eff}}
\newcommand{\Msun}{M_{\odot}}
\newcommand{\dd}{\mathrm{d}}
\usepackage{orcidlink}

\begin{document}

\title{Periodic orbits and gravitational wave signatures from magnetic dipoles around magnetized Kerr black holes}

\author{Shokhzod Jumaniyozov \orcidlink{0009-0009-6254-5608}}
\email{sh.jumaniyozov@newuu.uz}
\affiliation{New Uzbekistan University, Movarounnahr Street 1, Tashkent 100000, Uzbekistan}
\affiliation{Tashkent State Technical University, Tashkent 100095, Uzbekistan}

\author{Javlon~Rayimbaev\orcidlink{0000-0001-9293-1838}}
\email{javlon@astrin.uz}
\affiliation{Kimyo International University in Tashkent, Shota Rustaveli Street 156, Tashkent 100121, Uzbekistan}
\affiliation{University of Tashkent for Applied Sciences, Str. Gavhar 1, Tashkent 100149, Uzbekistan}

\author{Chengxun~Yuan~\orcidlink{0000-0002-2308-6703}}
\email{corresponding author: yuancx@hit.edu.cn}
\affiliation{School of Physics, Harbin Institute of Technology, Harbin 150001, People’s Republic of China}

\author{Satimbay~Palvanov \orcidlink{0000-0002-2694-6545}}
\email{satimbay@yandex.ru}
\affiliation{National University of Uzbekistan, Tashkent 100174, Uzbekistan}

\author{Khurshida Khaknazarova \orcidlink{0009-0009-1451-5199 }} \email{haknazarovahursida@gmail.com}
\affiliation{Samarkand State Pedagogical Institute, Spitamen Shokh Street 166, Samarkand 140100, Uzbekistan}

\author{Faisal~Javed\orcidlink{0000-0001-6970-1305}}
\email{faisaljaved.math@gmail.com}
\affiliation{School of Mathematical and Statistics, Zaozhuang University, Zaozhuang 277160, China}
\affiliation{Research Center of Astrophysics and Cosmology, Khazar University, Baku, AZ1096, 41 Mehseti Street, Azerbaijan}
\affiliation{Department of Mathematics, School of Science, University of Management and Technology,  Lahore, 54000, Pakistan}

\begin{abstract}
We study periodic orbits and the associated gravitational radiation of a magnetized (uncharged) test particle carrying a magnetic dipole moment with coupling constant $\beta$, moving in the equatorial plane of a rotating, magnetized Kerr black hole immersed in an external asymptotically uniform magnetic field, starting from the effective potential derived for such particles. We compute the marginally bound orbit (MBO) and the innermost stable circular orbit (ISCO) as functions of the black hole spin $a$ and the magnetic coupling $\beta$, and map out the allowed region of the orbital energy-angular momentum $(L, E)$ plane for bound motion. We then classify periodic orbits using the topological zoom-whirl scheme of Levin and Perez-Giz, characterized by three integers $(z,w,v)$ through the rational rotation number $q=w+v/z$, and construct a family of closed rosette orbits at fixed angular momentum. Using the numerical-kludge, restricted-quadrupole approximation for an extreme-mass-ratio inspiral consisting of a stellar-mass magnetized secondary orbiting a supermassive magnetized Kerr black hole, we compute the time-domain gravitational waveforms $h_+(t)$, $h_\times(t)$ produced by these periodic orbits and their frequency-domain characteristic strain, and compare the latter with the anticipated instrumental sensitivity curves of LISA, Taiji and TianQin. We find that the magnetic coupling $\beta$ systematically shifts the MBO and ISCO outward and lowers their orbital energy and angular momentum, that the zoom-whirl structure of the periodic orbits is imprinted directly on the burst-like morphology of the emitted waveform, and that the resulting gravitational-wave signals fall within the sensitivity band of upcoming space-based detectors for suitably close and massive sources.
\end{abstract}

\maketitle

\section{Introduction}

Black holes (BHs) are among the most striking predictions of general relativity (GR) and remain unrivalled laboratories for probing gravity in the strong-field regime. The direct detection of gravitational waves (GWs) from compact object mergers by the LIGO-Virgo-KAGRA collaboration~\cite{Abbott2016}, together with the Event Horizon Telescope images of the shadows of M87$^\ast$ and Sgr~A$^\ast$~\cite{EHT2019,EHTSgrA}, have opened an observational era in which the predictions of GR and of possible deviations from it can be tested directly in the vicinity of a horizon.
A particularly promising observational channel for future space-based GW detectors such as LISA~\cite{LISA2017}, Taiji~\cite{Taiji2020a,Taiji2020b} and TianQin~\cite{TianQin2016} is the extreme-mass-ratio inspiral (EMRI): the slow, adiabatic inspiral of a stellar-mass compact object into a supermassive BH (SMBH). Because the orbital evolution proceeds on a timescale much longer than a single orbital period, the trajectory of the small body can, over a limited number of cycles, be well approximated by a geodesic orbit of fixed energy and angular momentum. Such orbits often possess a rational ratio of their fundamental radial and azimuthal frequencies, $\Omega_r/\Omega_\phi \in \mathbb{Q}$, and therefore close on themselves after a finite number of radial librations: these are the periodic orbits of the underlying spacetime. Following Levin and Perez-Giz~\cite{LevinPerezGiz2008}, every periodic orbit in a stationary, axisymmetric, asymptotically flat spacetime can be labelled by three integers, the zoom number $z$, the whirl number $w$, and the vertex number $v$, through the rational rotation number
\begin{equation}
q \;=\; \frac{\Omega_\phi}{\Omega_r}-1 \;=\; w+\frac{v}{z}\, .
\end{equation}
Periodic orbits, and the characteristic ``zoom-whirl'' gravitational waveforms they produce, have since been studied extensively in Schwarzschild and Kerr spacetimes~\cite{LevinPerezGiz2008b,Glampedakis2002} and, more recently, in a wide variety of modified and alternative BH backgrounds (see, e.g.,~\cite{Lin2020,Deng2020,Zhang2020,Lin2021}), because periodic orbits provide a computationally efficient and physically transparent probe of how a given spacetime departs from the Kerr solution.

Astrophysical BHs are not expected to be magnetically neutral: accretion disks and magnetospheres readily thread a BH with an ordered, approximately uniform magnetic field, described in the test-field (Wald) approximation as living on a fixed Kerr background~\cite{Wald1974}. A compact object endowed with a magnetic dipole moment e.g., a magnetized neutron star or a strongly magnetized compact remnant orbiting such a magnetized BH experiences, in addition to the pure gravitational force, a magnetic dipole-field interaction that can be encoded, for an uncharged particle, in a single coupling constant $\beta$ and an interaction function $F(r)$ that modifies the effective potential of circular motion. The dynamics, collisional processes, and innermost stable and marginally bound orbits of charged and magnetized particles around magnetized rotating (Kerr) BHs were derived by Jumaniyozov, Khan, Rayimbaev, Abdujabbarov and Ahmedov~\cite{Jumaniyozov:2024hlg}, building on the broader literature on charged particle motion and magnetized compact objects in curved spacetimes~\cite{Aliev1989,Rayimbaev2020}. The present work specializes that formalism to the purely magnetized equatorial-plane case to study its periodic-orbit structure and gravitational-wave phenomenology.

The paper is organized as follows. Section~\ref{sec:spacetime} reviews the equatorial plane magnetized Kerr spacetime, the magnetic dipole interaction function $F(r)$, and the resulting effective potential for a magnetized test particle. Section~\ref{sec:isco-mbo} determines the MBO and ISCO and the allowed $(L, E)$ parameter space of bound orbits, as functions of the spin $a$ and the magnetic coupling $\beta$. Section~\ref{sec:periodic} constructs the periodic (zoom-whirl) orbits of the system, characterized by $(z,w,v)$, and tabulates their energies, radii, and orbital periods. Section~\ref{sec:gw} computes the time- and frequency-domain gravitational waveforms radiated by these periodic orbits within the restricted-quadrupole, numerical-kludge approximation, for an EMRI consisting of a magnetized stellar-mass secondary orbiting a Sgr~A$^\ast$-like magnetized supermassive BH, and compares the resulting characteristic strain with the projected sensitivities of LISA, Taiji, and TianQin. Section~\ref{sec:conclusions} summarizes our conclusions. Throughout, we use geometrized units $G=c=M=1$, with the BH mass $M$ setting the length and time scale unless stated otherwise.

\section{Spacetime and dynamics of a magnetized test particle}
\label{sec:spacetime}
\subsection{Rotating Kerr metric}
\label{sec:magnetized-kerr}

We consider a rotating BH of mass $M$ and spin parameter $a$, described by the vacuum Kerr solution of general relativity, and immersed in an external, asymptotically uniform magnetic field aligned with the black-hole rotation axis. Because astrophysical magnetic fields threading a BH are many orders of magnitude too weak to curve the surrounding spacetime appreciably, it is standard practice to treat such a field as a \emph{test field} superposed on the fixed Kerr background, following the classic construction of Wald~\cite{Wald1974}; we adopt this test-field approximation throughout. The BH itself therefore carries no magnetic charge, in accordance with the no-hair theorem, and the geometry is exactly that of the vacuum Kerr metric.

In Boyer--Lindquist coordinates $(t,r,\theta,\phi)$, the Kerr line element takes the stationary, axisymmetric form
\begin{equation}
ds^2 = g_{tt}\, dt^2 + g_{rr}\, dr^2 + g_{\theta\theta}\, d\theta^2 + g_{\phi\phi}\, d\phi^2 + 2 g_{t\phi}\, dt\, d\phi,
\label{eq:line_element}
\end{equation}
where the absence of $dt\,d\theta$ and $d\phi\,d\theta$ cross terms reflects the reflection symmetry of the metric about the equatorial plane, and the nonvanishing $g_{t\phi}$ component encodes the frame-dragging induced by the BH's rotation. The metric components entering Eq.~\eqref{eq:line_element} are
\begin{align}
g_{tt} &= -\left(1 - \frac{2Mr}{\Sigma}\right),
\qquad
g_{t\phi} = -\frac{2M a r \sin^2\theta}{\Sigma},
\qquad
g_{rr} = \frac{\Sigma}{\Delta},
\nonumber \\[4pt]
g_{\theta\theta} &= \Sigma,
\qquad
g_{\phi\phi} = \left(\rho^2 + \frac{2Mr a^2 \sin^2\theta}{\Sigma}\right) \sin^2\theta,
\label{eq:gphiphi}
\end{align}
with the auxiliary radial and polar functions
\begin{equation}
\Sigma = r^2 + a^2 \cos^2\theta,
\qquad
\Delta = r^2 - 2Mr + a^2,
\qquad
\rho^2 = r^2 + a^2 .
\label{eq:rhosq}
\end{equation}
The function $\Delta(r)$ controls the location of the horizons: its two roots,
\begin{equation}
r_{\rm eh}^{\pm} = M \pm \sqrt{M^2 - a^2},
\label{eq:horizon}
\end{equation}
give the outer (event) and inner (Cauchy) horizons, respectively, and coincide, $r_{\rm eh}^{+}=r_{\rm eh}^{-}=M$, in the extremal limit $a\to M$; throughout this work we restrict attention to sub-extremal BHs, $0\le a<M$, for which $r_{\rm eh}^{+}$ is real and the exterior spacetime $r>r_{\rm eh}^{+}$ is free of coordinate or curvature pathologies. Setting $a=0$ in Eqs.~\eqref{eq:gphiphi}--\eqref{eq:rhosq} recovers the familiar Schwarzschild metric, while $M\to0$ recovers flat spacetime in oblate spheroidal coordinates, providing two useful consistency checks on the formalism developed below.
Having fixed the background geometry, we next specify the external magnetic field threading it and, in Sec.~\ref{sec:wald-field}, derive the ZAMO-frame field components that will source the dipole interaction energy of a magnetized test particle.

\subsection{The Wald magnetic field}\label{sec:wald-field}
The construction rests on a simple but powerful fact: since the Kerr metric is a vacuum (Ricci-flat) solution, every Killing vector $\xi_\mu$ automatically satisfies the source-free Maxwell equation $\nabla^\nu\nabla_\nu\xi_\mu=-R_{\mu\nu}\xi^\nu=0$ once it is identified with a vector potential $A_\mu=\xi_\mu$. Wald~\cite{Wald1974} exploited exactly this to construct the unique stationary, axisymmetric, asymptotically uniform test magnetic field of strength $B$ that is aligned with the black-hole spin axis and leaves the horizon uncharged,
\begin{equation}
A_\mu \;=\; \frac{B}{2}\Big(\xi^{(\phi)}_\mu + 2a\,\xi^{(t)}_\mu\Big) ,
\qquad \xi^{(t)}=\partial_t,\ \ \xi^{(\phi)}=\partial_\phi ,
\label{eq:wald}
\end{equation}
built as a combination of the two Killing vectors generated by time translation and axial rotation. In the equatorial plane this potential has components $A_t=\tfrac{B}{2}(g_{t\phi}+2a\,g_{tt})$ and $A_\phi=\tfrac{B}{2}(g_{\phi\phi}+2a\,g_{t\phi})$; setting $a=0$ recovers the familiar flat-space-like form $A_\phi=\tfrac12 Br^2\sin^2\theta$, as it should~\cite{Oteev:2025csb}. From $A_\mu$ the field strength $F_{\mu\nu}=\partial_\mu A_\nu-\partial_\nu A_\mu$ follows directly, and in the equatorial plane it reduces to only two nonzero, purely radial components, $F_{tr}=\partial_r A_t$ and $F_{\phi r}=\partial_r A_\phi$.

\subsection{Dipole coupling of an uncharged particle} 
A test particle carrying zero electric charge but a magnetic dipole moment $\mu$ does not feel the Lorentz force; instead, following the covariant polarization-tensor formalism of Refs.~\cite{Tursunov16,Preti04}, its interaction with the external field is governed by the scalar
\begin{equation}
2U \;=\; D^{\mu\nu}F_{\mu\nu} ,
\label{eq:U-def}
\end{equation}
where $F_{\mu\nu}$ is the electromagnetic field tensor and $D^{\mu\nu}$ is the polarization tensor built from the particle's magnetic dipole moment 4-vector $\mu^\nu$ and 4-velocity $u^\nu$,
\begin{equation}
D^{\mu\nu} \;=\; \eta^{\mu\nu\sigma\rho}u_\sigma\mu_\rho ,
\qquad D^{\mu\nu}u_\nu=0 ,
\label{eq:Dtensor}
\end{equation}
with $\eta_{\mu\nu\sigma\rho}=\sqrt{-g}\,\epsilon_{\mu\nu\sigma\rho}$, $\eta^{\mu\nu\sigma\rho}=-\tfrac{1}{\sqrt{-g}}\epsilon^{\mu\nu\sigma\rho}$ the covariant Levi-Civita tensor built from the totally antisymmetric symbol $\epsilon_{\mu\nu\sigma\rho}$ and $g=\det(g_{\mu\nu})$. Splitting $F_{\mu\nu}$ into electric and magnetic parts as measured by an observer of 4-velocity $u^\mu$,
\begin{equation}
F_{\mu\nu} \;=\; 2u_{[\mu}E_{\nu]}+\eta_{\mu\nu\sigma\rho}u^\sigma B^\rho ,
\qquad
B^\mu \;=\; \tfrac12\,\eta^{\mu\nu\sigma\rho}F_{\nu\sigma}u_\rho ,
\label{eq:Bfield}
\end{equation}
gives the magnetic field 4-vector felt by that observer. Taking the observer to be a zero-angular-momentum observer (ZAMO) and orienting the dipole moment perpendicular to the equatorial plane, parallel to the field lines, $\mu_{\hat\imath}=(0,\mu_{\hat\theta},0)$, reduces the interaction scalar to $U=2\mu_{\hat\theta}B^{\hat\theta}$, with the ZAMO-frame field component sourced by the Wald potential~\eqref{eq:wald} given by~\cite{Tursunov16}

\begin{eqnarray}
B^{\hat\theta} = \frac{B\sqrt{g_{\theta\theta}}}{\sqrt{-g}}
\left[
a\left(
u_{\phi}\, g_{tt,r}
-
u_{t}\, g_{t\phi,r}
\right)
+\frac{1}{2}
\left(
u_{\phi}\, g_{t\phi,r}
-
u_{t}\, g_{\phi\phi,r}
\right)
\right],
\label{eq:Btheta}
\end{eqnarray}
four velocity components are as follows $u^{\mu}_{ZAMO}=\{u^t,0,0,u^{\phi}\}$, (where  $(u^t)^2=g_{\phi \phi}/(g_{t\phi}^2-g_{tt} g_{\phi \phi})$,  $u^{\phi }=-g_{t \phi}u^t/g_{\phi \phi }$). Evaluated on the equatorial plane, Eq.~\eqref{eq:Btheta} yields an interaction energy of exactly the form
\begin{equation}
U(r) \;=\; 2\mu_{\hat\theta}B^{\hat\theta} \;=\; 2\mu B\,F(r) ,
\label{eq:U-equatorial}
\end{equation}
i.e., a position-dependent energy proportional to a dimensionless radial profile function $F(r)$ in the equatorial plane as follows
\begin{equation}
F(r) \;=\; -\,\frac{a^2+r^3}{r^3\sqrt{\,1+\dfrac{2}{r}+\dfrac{4}{a^2+(r-2)r}\,}} \; ,
\label{eq:Fr}
\end{equation}
This closed form satisfies $F(r)\to -1$ as $r\to\infty$, so that a magnetized particle released at rest at infinity has specific energy $E_{\rm esc}=|1-\beta|$ rather than unity, a generalization of the usual asymptotic (``escape'') energy condition used throughout this work to define the MBO. We fix the fiducial working point $a=0.3M$, $\beta=0.1$ for the numerical illustrations below (varying both parameters where indicated), consistent with the values adopted in the accompanying numerical implementation.

Before proceeding, it is instructive to translate the coupling constant $\beta$ into the astrophysical parameters of a realistic secondary, so as to assess whether the range $\beta\in[0,\,0.30]$ explored throughout this work corresponds to a physically motivated regime. Modelling the magnetized secondary as a uniformly magnetized sphere, its magnetic dipole moment is $\mu_{\rm NS}=\tfrac{1}{2}B_{\rm NS}R_{\rm NS}^3$, where $B_{\rm NS}$ and $R_{\rm NS}$ are the surface magnetic field and radius of the neutron star. Combining this with the definition $\beta=2\mu B/m$ and restoring physical units gives the practical estimate~\cite{Narzilloev2021}
\begin{equation}
\beta \;\simeq\; \frac{11}{250}
\left(\frac{B_{\rm NS}}{10^{12}\,\mathrm G}\right)
\left(\frac{R_{\rm NS}}{10^{6}\,\mathrm{cm}}\right)^{3}
\left(\frac{B_{\rm ext}}{10\,\mathrm G}\right)
\left(\frac{m_{\rm NS}}{M_\odot}\right)^{-1},
\label{eq:beta-astro}
\end{equation}
where $B_{\rm ext}$ is the strength of the external (Wald) field at the location of the secondary.
 
For an ordinary, non-magnetar neutron star with a canonical surface field $B_{\rm NS}\sim10^{12}$--$10^{13}\,$G, radius $R_{\rm NS}\simeq10\,$km, and mass $m_{\rm NS}\simeq1.4\, M_\odot$, orbiting a supermassive black hole threaded by a modest external field $B_{\rm ext}\sim10$--$10^{2}\,$G — the range typically adopted for the magnetosphere of a low-luminosity SMBH such as Sgr~A$^\ast$~\cite{Tursunov16} — Eq.~\eqref{eq:beta-astro} gives $\beta\sim0.03$--$0.4$. This range brackets both the fiducial value $\beta=0.1$ adopted for the numerical illustrations of Secs.~III--V and the upper value $\beta=0.30$ reached in the scans of Sec.~IV\, D, confirming that the couplings explored in this work correspond to an ordinary, rather than an extreme, magnetized neutron-star companion. For reference, the only magnetar presently known to orbit a supermassive black hole, PSR~J1745$-$2900 near Sgr~A$^\ast$, with dipole moment $\mu\simeq1.6\times10^{32}\,\mathrm{G\,cm^3}$~\cite{Mori2013}, would correspond to $\beta\simeq0.7\,(B_{\rm ext}/10\,\mathrm G)$~\cite{Narzilloev2021} — several times larger than the upper bound $\beta=0.30$ considered here even at the low end of $B_{\rm ext}$. The interval $\beta\in[0,\,0.30]$ adopted throughout this paper is therefore a conservative, astrophysically motivated choice, safely below the magnetar-class regime.

The corresponding covariant equation of motion follows from the effective Lagrangian for a particle with electric charge $q$ and magnetic dipole moment $\mu$ \cite{UktamjonUktamov:2025aqz},
\begin{equation}
\mathcal L \;=\; \tfrac12\Big[(m+U)\,g_{\mu\nu}u^\mu u^\nu - kU\Big] + q A_\mu u^\mu ,
\label{eq:preti-lagrangian}
\end{equation}
with $k$ an order-unity constant fixed by the pole-dipole equations of motion. For the uncharged particle considered here, $q=0$, and only the first term survives; since $U(r)$ carries no velocity dependence, the conjugate momenta following from Eq.~\eqref{eq:preti-lagrangian} are $p_\mu=(m+U)g_{\mu\nu}u^\nu$, so that the timelike normalization $g_{\mu\nu}u^\mu u^\nu=-1$ gives directly
\begin{equation}
g^{\mu\nu}p_\mu p_\nu \;=\; -\,\big(m+U\big)^2 ,
\end{equation}
which, upon identifying $U/m=\beta F(r)$ with $\beta\equiv2\mu B/m$ as anticipated in Eq.~\eqref{eq:U-equatorial} above, is precisely the generalized mass-shell condition \cite{UktamjonUktamov:2025aqz}
\begin{equation}
g^{\mu\nu}p_\mu p_\nu \;=\; -\,m^2\big(1+\beta F(r)\big)^2 ,
\label{eq:massshell}
\end{equation}
used throughout this work in place of the usual $g^{\mu\nu}p_\mu p_\nu=-m^2$.  Carrying out the ZAMO frame projection of the field~\eqref{eq:wald} explicitly in the Kerr background thus yields the closed-form radial profile $F(r)$ given above in Eq.~\eqref{eq:Fr}.

\subsection{Effective potential}
\label{sec:HJ}
To make explicit how the dipole term $(1+\beta F(r))^2$ used above arises dynamically rather than being postulated, we briefly rederive it from the Hamilton-Jacobi equation for an uncharged, magnetized test particle moving on the fixed (test-field) Kerr background, following the Wald construction for the external magnetic field~\cite{Wald1974} and the dipole-coupling reduction of Ref.~\cite{Jumaniyozov:2024hlg}.

Because the right-hand side of Eq.~\eqref{eq:massshell} depends only on $r$ in the equatorial plane, the modified mass-shell condition remains separable in the same way as the ordinary geodesic ($\beta=0$) case. Writing the Hamilton-Jacobi equation 

\begin{equation}
    g^{\mu\nu}\,\frac{\partial \mathcal{S}}{\partial x^\mu}\,\frac{\partial \mathcal{S}}{\partial x^\nu} = -m^2\left(1+\frac{U}{m}\right)^2 ,
\end{equation}
with the standard separated ansatz
$S(t,r,\phi) = -Et + L\phi + S_r(r)$ and substituting the metric components together with $p_t$, $p_\phi$, and $p_r=\dd S_r/\dd r$, gives
\begin{equation}
g^{tt}E^2 - 2g^{t\phi}EL + g^{\phi\phi}L^2 + g^{rr}\Big(\frac{\dd S_r}{\dd r}\Big)^{2} = -m^2\big(1+\beta F(r)\big)^2 .
\label{eq:HJ-explicit}
\end{equation}
With $\alpha(r)=-g^{tt}(r)$, $\delta(r,L)=g^{t\phi}(r)L$, $\gamma(r,L)=g^{\phi\phi}(r)L^2+(1+\beta F(r))^2$, Eq.~\eqref{eq:HJ-explicit} rearranges to $g^{rr}(S_r')^2=\alpha E^2+2\delta E-\gamma$; using $\dot r=p^r=g^{rr}S_r'$ and $g^{rr}=\Delta(r)/r^2$ then gives directly
\begin{equation}
\dot r^2(r,E,L) = \frac{\Delta(r)}{r^2}\Big[\alpha(r)E^2+2\delta(r,L)E-\gamma(r,L)\Big] ,
\label{eq:HJ-radial}
\end{equation}
where an overdot denotes differentiation with respect to the affine parameter $\lambda$.
To study the radial motion, we define an effective potential
$V_{\rm eff}$ such that:
\begin{equation}
\left(\frac{dr}{d\lambda}\right)^2 + V_{\rm eff} = E.
\label{eq:68}
\end{equation}

The condition $\dot r=0$ at fixed $L$ defines the effective potential $V_\eff(r, L)$ for circular motion, obtained as the physical (outer) root of the quadratic $\alpha E^2+2\delta E-\gamma=0$:
\begin{equation}
V_\eff(r,L) \;=\; \frac{-\delta(r,L)+\sqrt{\delta(r,L)^2+\alpha(r)\gamma(r,L)}}{\alpha(r)}\, ,
\label{eq:Veff}
\end{equation}
The sign of the discriminant root is fixed so that Eq.~\eqref{eq:Veff} reduces to the standard Kerr geodesic result when $\beta=0$; we have verified this reduction numerically against the Bardeen-Press-Teukolsky Kerr ISCO formula~\cite{Bardeen1972} for $a\neq0$ and against $r_{\rm ISCO}=6M$, $r_{\rm MBO}=4M$ at $a=0$. Bound motion between a periastron $r_1$ and an apastron $r_2$ requires $E<V_\eff(r,L)$ at those turning points and $\dot r^2\ge0$ in between, i.e., $r_1,r_2$ are adjacent roots of $\dot r^2(r,E,L)=0$.

\begin{figure}[t]
\centering
\includegraphics[width=0.98\linewidth]{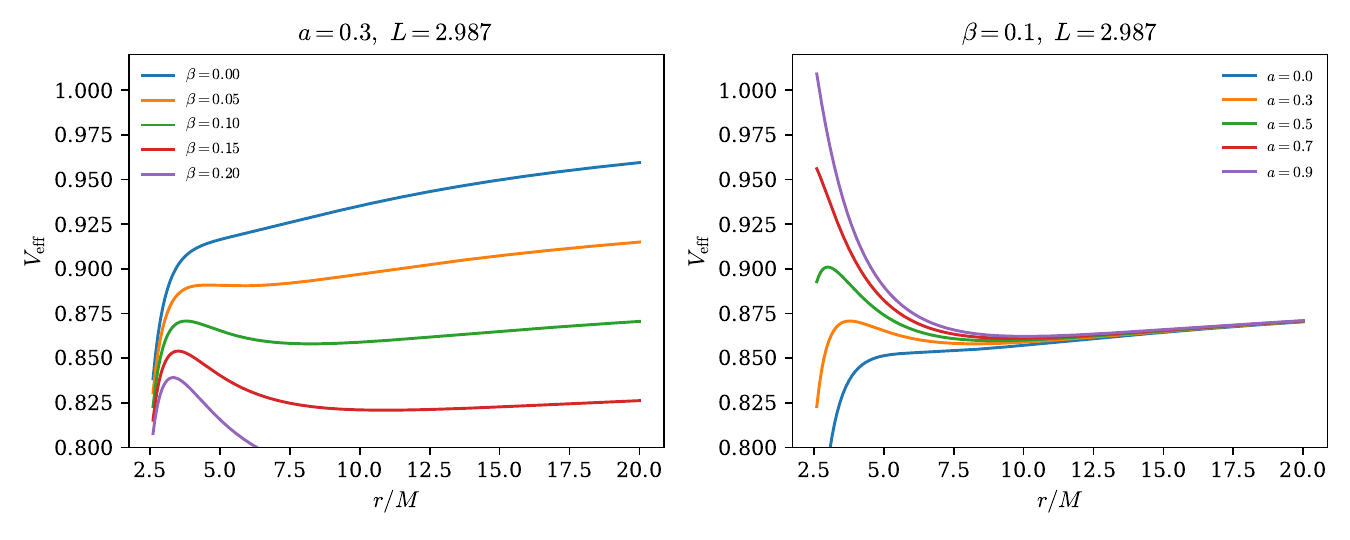}
\caption{Radial dependence of the effective potential $V_\eff(r,L)$ of a magnetized test particle at fixed angular momentum $L=\tfrac12(L_{\rm ISCO}+L_{\rm MBO})$ (evaluated at the fiducial point $a=0.3$, $\beta=0.1$, and held fixed while $\beta$ or $a$ is varied for illustration). Left: dependence on the magnetic coupling $\beta$ at fixed spin $a=0.3$. Right: dependence on the spin $a$ at fixed $\beta=0.1$.}
\label{fig:veff}
\end{figure}

Figure~\ref{fig:veff} shows the radial dependence of $V_\eff(r,L)$ at fixed $L=\tfrac12(L_{\rm ISCO}+L_{\rm MBO})$ for several values of the magnetic coupling $\beta$ (left panel, at fixed spin $a=0.3$) and of the spin $a$ (right panel, at fixed $\beta=0.1$). Increasing $\beta$ lowers the potential barrier and shifts its peak inward, while increasing $a$ raises the potential near the horizon, mirroring the familiar frame-dragging enhancement of the Kerr effective potential.

\section{Marginally bound and innermost stable circular orbits}
\label{sec:isco-mbo}

Circular orbits of magnetized particles satisfy $\partial V_\eff/\partial r=0$ at fixed $L$. The innermost stable circular orbit (ISCO) additionally requires the orbit to sit at the inflection point of the potential~\cite{Alloqulov:2025bxh,Jumaniyozov:2026gxv},
\begin{equation}
\frac{\partial V_\eff}{\partial r}=0, \qquad \frac{\partial^2 V_\eff}{\partial r^2}=0,
\label{eq:isco-cond}
\end{equation}
while the marginally bound orbit (MBO) is the circular orbit whose energy equals the generalized escape energy $E_{\rm esc}=|1-\beta|$ introduced above,
\begin{equation}
V_\eff(r,L)=E_{\rm esc}, \qquad \frac{\partial V_\eff}{\partial r}=0.
\label{eq:mbo-cond}
\end{equation}

\begin{figure}[t]
\centering
\includegraphics[width=0.9\linewidth]{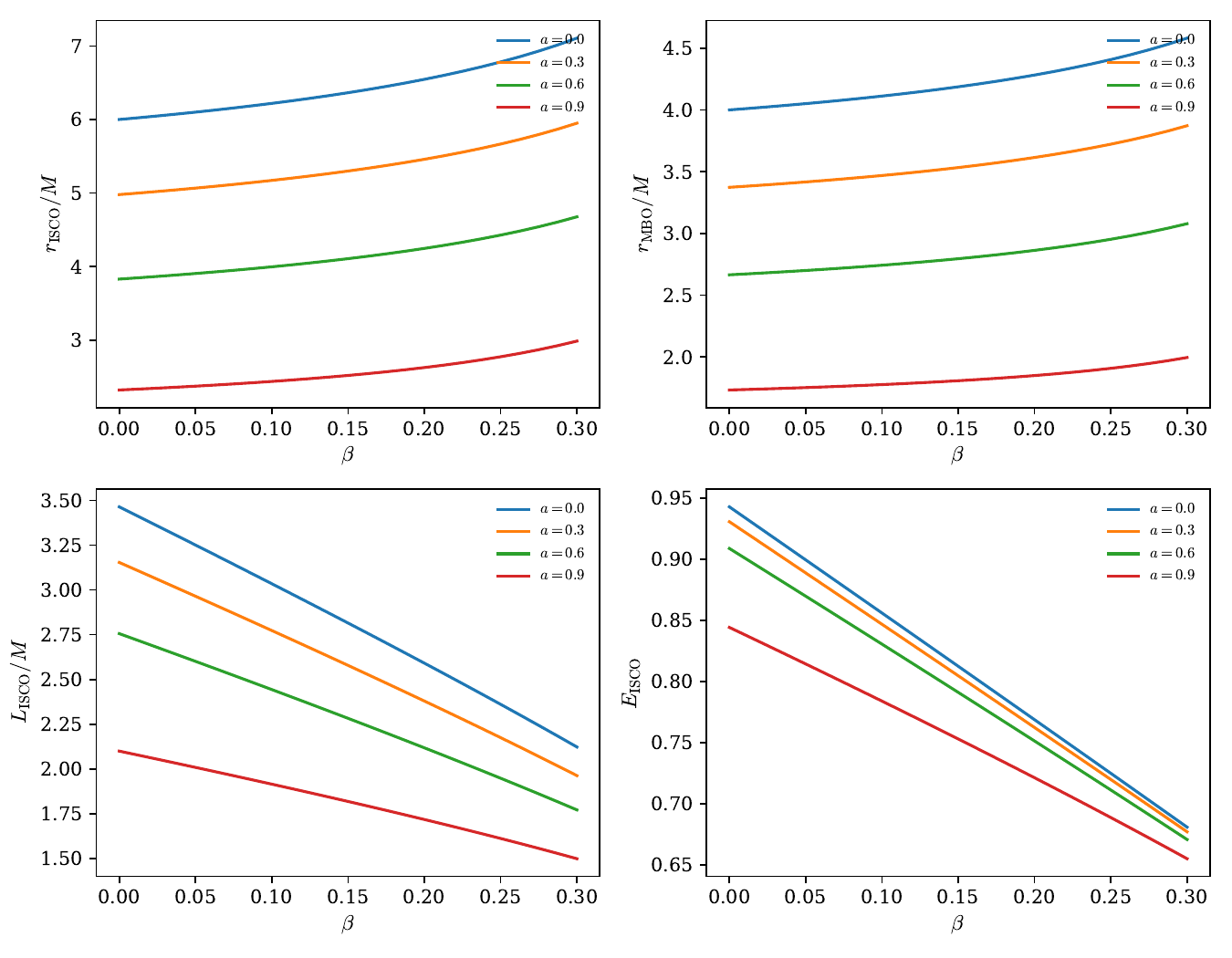}
\caption{Dependence of the ISCO radius, MBO radius, ISCO angular momentum and ISCO energy on the magnetic coupling $\beta$, for spins $a=0,0.3,0.6,0.9$.}
\label{fig:isco-mbo}
\end{figure}

\label{sec:methods}
At the fiducial point $a=0.3M$, $\beta=0.1$, this procedure gives
\begin{align}
r_{\rm ISCO}&=5.17135\,M, & L_{\rm ISCO}&=2.77327\,M, & E_{\rm ISCO}&=0.84683, \\
r_{\rm MBO}&=3.46878\,M, & L_{\rm MBO}&=3.20004\,M, & E_{\rm MBO}&=0.90000\;(=1-\beta).
\end{align}
Table~\ref{tab:isco-mbo} lists $r_{\rm ISCO}$, $L_{\rm ISCO}$, $E_{\rm ISCO}$, $r_{\rm MBO}$, $L_{\rm MBO}$ and $E_{\rm MBO}$ as functions of $\beta$ at fixed spin $a=0.3M$; Fig.~\ref{fig:isco-mbo} extends this to several representative spins $a=0,0.3,0.6,0.9$. In all cases, the radii of both the ISCO and the MBO increase monotonically with $\beta$, while their orbital energy and angular momentum decrease: a stronger magnetic dipole coupling makes both orbits less tightly bound and pushes them further from the horizon, where increasing charge instead shrinks the ISCO and MBO. As expected, increasing the spin $a$ at fixed $\beta$ shrinks both radii through ordinary frame-dragging, exactly as in the pure Kerr case.

\begin{table}[t]
\centering
\caption{ISCO and MBO radius, angular momentum and energy as functions of the magnetic coupling $\beta$, at fixed spin $a=0.3M$. The $\beta=0.10$ row is the fiducial working point used throughout this paper and reproduces the independent \textsc{Mathematica} computation to 6 significant figures.}
\label{tab:isco-mbo}
\begin{tabular}{@{}ccccccc@{}}
\toprule
$\beta$ & $r_{\rm ISCO}/M$ & $L_{\rm ISCO}/M$ & $E_{\rm ISCO}$ & $r_{\rm MBO}/M$ & $L_{\rm MBO}/M$ & $E_{\rm MBO}$ \\
\midrule
0.00 & 4.9786 & 3.1536 & 0.9306 & 3.3733 & 3.6733 & 1.0000 \\
0.05 & 5.0666 & 2.9648 & 0.8888 & 3.4166 & 3.4380 & 0.9500 \\
0.10 & 5.1714 & 2.7733 & 0.8468 & 3.4688 & 3.2000 & 0.9000 \\
0.15 & 5.2989 & 2.5785 & 0.8047 & 3.5331 & 2.9588 & 0.8500 \\
0.20 & 5.4586 & 2.3795 & 0.7624 & 3.6148 & 2.7134 & 0.8000 \\
0.25 & 5.6660 & 2.1749 & 0.7199 & 3.7223 & 2.4623 & 0.7500 \\
0.30 & 5.9497 & 1.9628 & 0.6772 & 3.8719 & 2.2032 & 0.7000 \\
\bottomrule
\end{tabular}
\end{table}

\begin{figure}[t]
\centering
\includegraphics[width=0.72\linewidth]{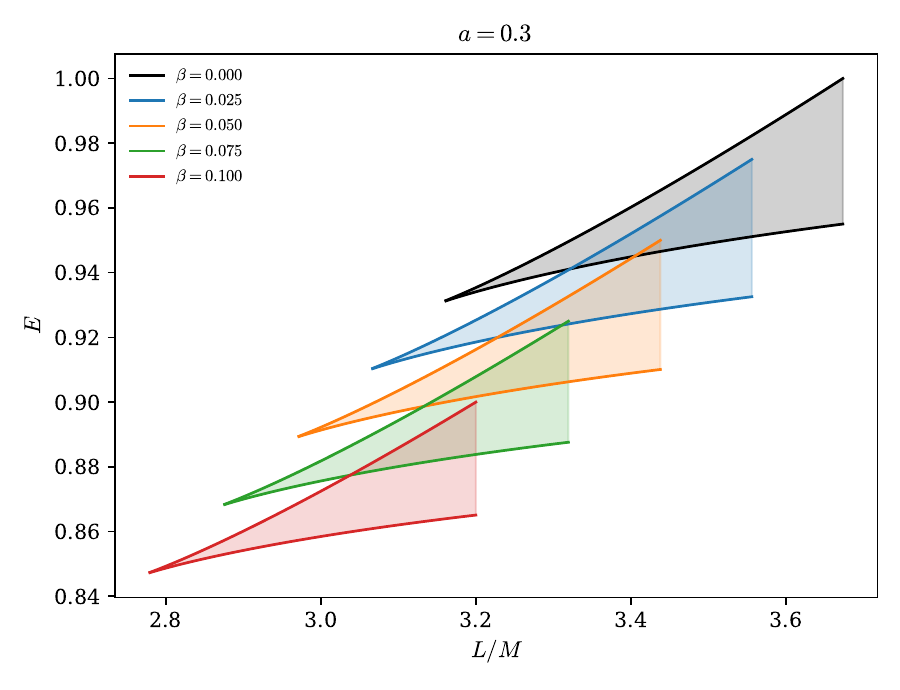}
\caption{Allowed region of the orbital angular momentum-energy $(L,E)$ plane for bound orbits of a magnetized particle around the Kerr BH, for $\beta=0,0.025,0.05,0.075,0.10$ at fixed spin $a=0.3$. The lower/upper boundary of each shaded region is the energy of the stable/unstable circular orbit at that $L$; the two boundaries meet at $L=L_{\rm ISCO}$.}
\label{fig:bound-region}
\end{figure}

For a bound orbit to exist at fixed $L$, the energy must lie between the energy of the (outer) stable circular orbit and that of the (inner) unstable circular orbit at the same $L$; scanning $L$ between $L_{\rm ISCO}$ and $L_{\rm MBO}$ and solving $\partial V_\eff/\partial r=0$ for both branches (again by continuation in $L$) traces out the boundary of the allowed $(L,E)$ region for bound motion, shown in Fig.~\ref{fig:bound-region} for several values of $\beta$ at $a=0.3M$. As $\beta$ increases, this region shifts toward lower $L$ and lower $E$, consistent with the trend already seen in Table~\ref{tab:isco-mbo}.

As a concrete illustration used later to seed the periodic-orbit construction, we fix $L=L_{\rm ISCO}+\varepsilon\,(L_{\rm MBO}-L_{\rm ISCO})$ with $\varepsilon=0.5$, giving $L=2.98666\,M$ at the fiducial point. The corresponding unstable and stable circular orbits are located at
\begin{equation}
(r_{\rm unstable}, E_{\rm unstable}) = (3.78387\,M,\ 0.87074), \qquad
(r_{\rm stable}, E_{\rm stable}) = (8.24501\,M,\ 0.85794),
\end{equation}
and, choosing an illustrative bound-orbit energy $E=E_{\rm stable}+0.6\,(E_{\rm unstable}-E_{\rm stable})=0.86562$, the two roots of $\dot r^2(r,E,L)=0$ nearest the stable branch give the periastron and apastron,
\begin{equation}
r_{\rm peri} = 4.93781\,M, \qquad r_{\rm apo} = 15.57574\,M .
\end{equation}
These periastron/apastron finder routines, seeded from the previous solution at each step, are used pervasively in the remainder of the paper.

\section{Periodic orbits around the magnetized Kerr BHs}
\label{sec:periodic}

\subsection{Zoom-whirl classification and the rational rotation number}

\begin{figure}[t]
\centering
\includegraphics[width=0.98\linewidth]{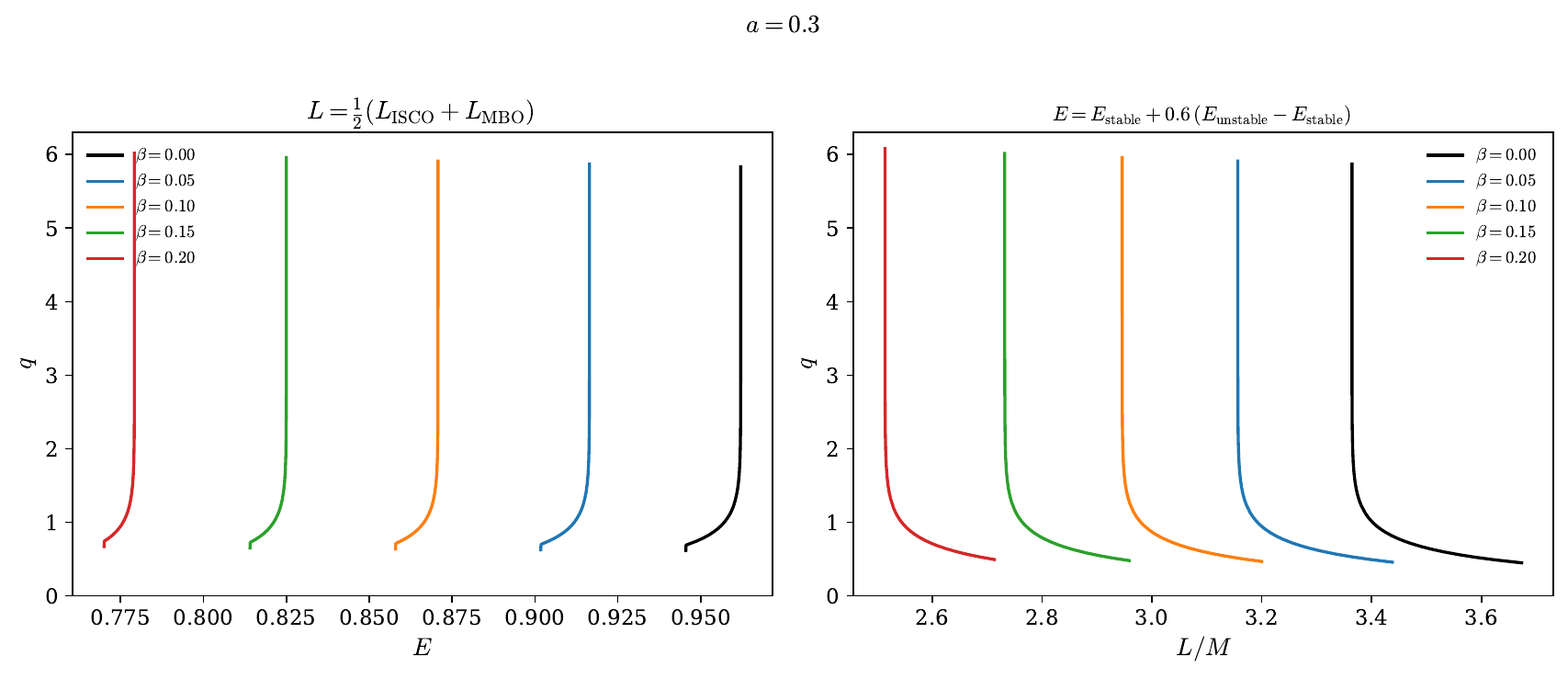}
\caption{Rational rotation number $q$ as a function of energy at fixed $L=\tfrac12(L_{\rm ISCO}+L_{\rm MBO})$ (left) and as a function of angular momentum at fixed $E=E_{\rm stable}+0.6(E_{\rm unstable}-E_{\rm stable})$ (right), spin $a=0.3$, for several values of the magnetic coupling $\beta$.}
\label{fig:qE}
\end{figure}

Every bound orbit oscillates between periastron $r_1$ and apastron $r_2$ while precessing in $\phi$. Using the coordinate-velocity functions obtained from the inverse $(t,\phi)$ metric block,
\begin{equation}
\dot\phi(r,E,L) = -g^{t\phi}(r)E+g^{\phi\phi}(r)L, \qquad
\dot t(r,E,L) = -g^{tt}(r)E+g^{t\phi}(r)L,
\end{equation}
the total azimuthal angle accumulated over one full radial libration (periastron $\to$ apastron $\to$ periastron) is~\cite{Wang:2025hla,Choudhury:2025qsh,Jumaniyozov:2026jce,Lu:2025cxx}
\begin{equation}
\Delta\phi(E,L) = 2\int_{r_1}^{r_2}\frac{\dot\phi(r,E,L)}{\sqrt{\dot r^2(r,E,L)}}\,\dd r ,
\end{equation}
and the rational rotation number, following Levin and Perez-Giz~\cite{LevinPerezGiz2008},
\begin{equation}
q(E,L) \;=\; \frac{\Delta\phi(E,L)}{2\pi}-1 \;=\; w+\frac{v}{z}\, ,
\label{eq:qEL}
\end{equation}
is an integer $w$ (the whirl number, counting extra near-circular loops the orbit makes close to periastron) plus a proper fraction $v/z$ with $0\le v<z$ (the zoom number $z$ being the number of radial librations, or ``leaves'', needed for the orbit to close). Orbits with irrational $q$ never close and instead trace a dense rosette that reveals the same underlying zoom-whirl skeleton. Figure~\ref{fig:qE} shows $q(E)$ at fixed $L=\tfrac12(L_{\rm ISCO}+L_{\rm MBO})$ (left panel) and $q(L)$ at fixed $E=E_{\rm stable}+0.6(E_{\rm unstable}-E_{\rm stable})$ (right panel, the same energy used to seed the periastron/apastron construction of Sec.~\ref{sec:isco-mbo}), for several values of $\beta$: $q$ increases monotonically with $E$ and diverges as $E\to E_{\rm unstable}$, where the orbit whirls indefinitely close to the unstable circular orbit, and correspondingly decreases with $L$ and diverges as $L$ approaches its lower bound at fixed $E$  exactly as found for the dyonic ModMax BH in Ref.~\cite{Alloqulov2026} and for Kerr geodesics more generally~\cite{LevinPerezGiz2008}. Increasing $\beta$ shifts both curves toward lower $E$ and lower $L$, consistent with the outward, lower-energy shift of the whole bound-orbit region already seen in Fig.~\ref{fig:bound-region}.

\subsection{Constructing rational orbit initial data}

Given target integers $(z,w,v)$ and a fixed angular momentum $L$, we invert the monotonic $q(E)$ curve of Sec.~\ref{sec:periodic} (via a cubic interpolant) to obtain the energy $E$ for which $q(E,L)=w+v/z$, and release the particle from rest at the corresponding apastron $r_0={\rm apastron}(E,L)$. The full state of a periodic orbit is then $\{t_0,r_0,\phi_0,E,L\}=\{0,r_0,0,E,L\}$, since $(E,L)$ already fix the orbit completely through the two Killing vectors. This is considerably simpler than the general Hamiltonian phase-space construction needed for a generic (non-conserved) system, and follows the equatorial-plane specialization of the static-spacetime rational-orbit algorithm used in Ref.~\cite{Alloqulov2026}.

Fixing $L=2.98666\,M$ (the same value used in Sec.~\ref{sec:isco-mbo}) at the fiducial point $a=0.3$, $\beta=0.1$, we construct six periodic orbits with $z=1,\dots,6$, all with whirl number $w=1$ and vertex number $v=z-1$ (i.e., $q=1+\frac{z-1}{z}$), plus one irrational orbit obtained by perturbing the $z=5$ rational number by $\delta q=1/(100\,z)=0.002$. Table~\ref{tab:orbits} lists the resulting $(z,w,v)$, rational number $q$, apastron radius $r_0$, orbital energy $E$, and the full closed-orbit period $T_{\rm orbit}=2z\, T_{\rm leaf}$, where $T_{\rm leaf}=\int_{r_1}^{r_2}\dd r/\sqrt{\dot r^2(r, E, L)}$ is the time (in affine-parameter units) for a single periastron-to-apastron half-libration.

\begin{table}[ht]
\centering
\caption{The seven periodic/quasi-periodic orbits constructed at fixed $L=2.98666\,M$, $a=0.3$, $\beta=0.1$. $T_{\rm orbit}$ is given in units of $M$ (affine-parameter/geometrized time).}
\label{tab:orbits}
\begin{tabular}{@{}cccccc@{}}
\toprule
$(z,w,v)$ & $q=w+v/z$ & $r_0/M$ (apastron) & $E$ & $T_{\rm orbit}/M$ \\
\midrule
$(1,1,0)$ & $1$        & $16.3493$ & $0.866554$ & $335.033$  \\
$(2,1,1)$ & $3/2$      & $19.5524$ & $0.870066$ & $806.987$  \\
$(3,1,2)$ & $5/3$      & $19.8544$ & $0.870369$ & $1240.575$ \\
$(4,1,3)$ & $7/4$      & $19.9498$ & $0.870463$ & $1670.064$ \\
$(5,1,4)$ & $9/5$      & $19.9968$ & $0.870509$ & $2098.892$ \\
$(6,1,5)$ & $11/6$     & $20.0223$ & $0.870534$ & $2526.765$ \\
\midrule
irrational, $(z_0,w_0,v_0){=}(5,1,4)+\delta q$ & $901/500=1.802$ & $19.9986$ & $0.870511$ & $\left(3\times2098.892\right)^\dagger$ \\
\bottomrule
\end{tabular}
\\[2pt]
{\footnotesize $^\dagger$Integrated for three times the base $z{=}5$ orbital period to reveal the denser, non-closing rosette structure.}
\end{table}

\subsection{Orbital shapes}

\begin{figure}[t]
\centering
\includegraphics[width=0.98\linewidth]{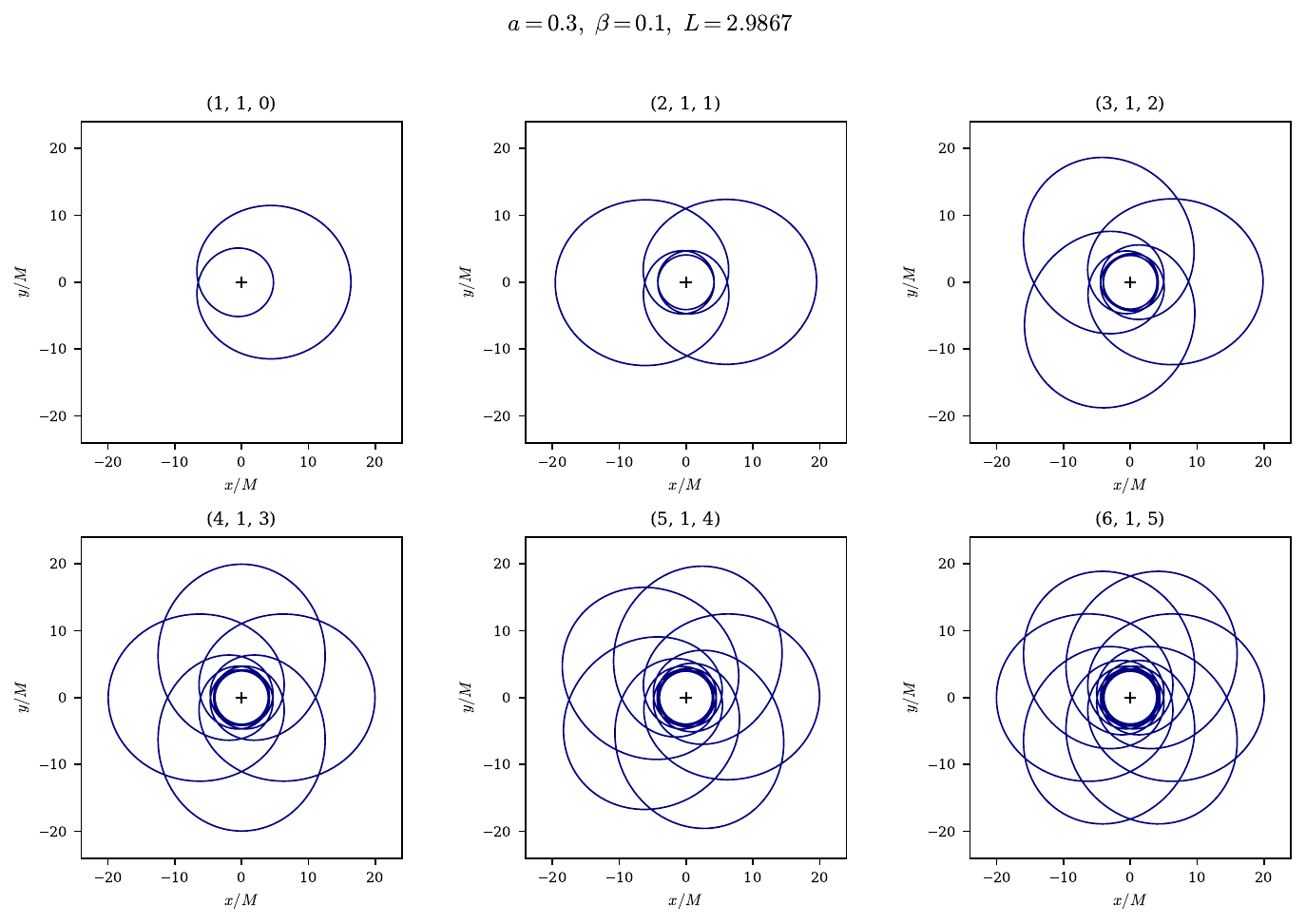}
\caption{Periodic (rosette) orbits for the six $(z,w,v)$ configurations of Table~\ref{tab:orbits}, at fixed $a=0.3$, $\beta=0.1$, $L=2.98666\,M$. The black cross marks the BH location.}
\label{fig:rosettes}
\end{figure}

\begin{figure}[t]
\centering
\includegraphics[width=0.42\linewidth]{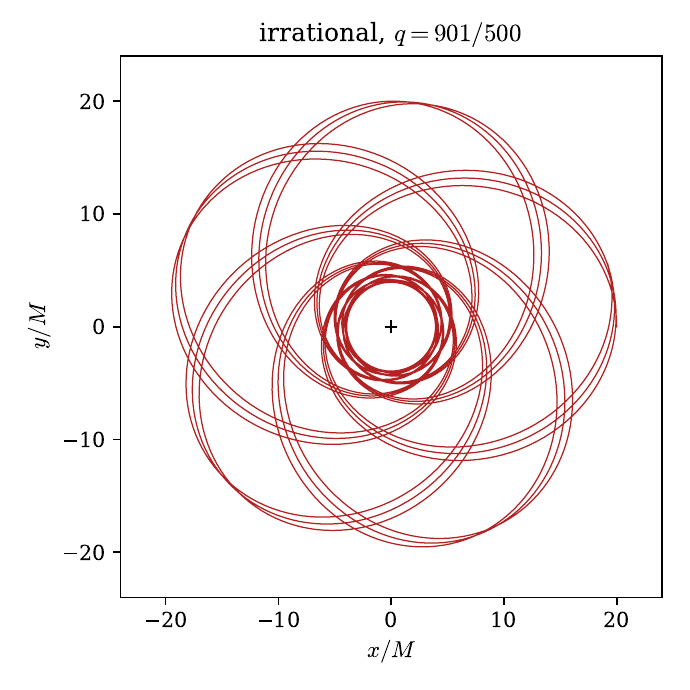}
\caption{Quasi-periodic orbit with irrational rotation number $q=901/500+\delta q\approx1.804$ (built by perturbing the $z=5$ rational orbit), integrated for three orbital periods.}
\label{fig:irrational}
\end{figure}

Given $(E,L,r_0)$ for each entry of Table~\ref{tab:orbits}, we integrate the orbit forward in the affine parameter $\lambda$. Rather than integrating the first-order radial constraint $\dot r=\pm\sqrt{\dot r^2(r,E,L)}$ directly, which is numerically delicate at the turning points where $\dot r^2\to0$, we differentiate Eq.~\eqref{eq:HJ-radial} once more to obtain a genuine second-order radial oscillator~\cite{Ahmed:2025azu,Sharipov:2025yfw,Zare:2025aek},
\begin{equation}
Q(r;E,L)\equiv\dot r^2(r,E,L),   \qquad  \ddot r(\lambda) = \tfrac12\,Q'\!\big(r(\lambda);E,L\big),
\end{equation}
integrated together with the first-order equations $\dot\phi=\dot\phi(r,E,L)$ and $\dot t=\dot t(r,E,L)$ from initial data $r(0)=r_0$, $\dot r(0)=0$, $\phi(0)=t(0)=0$, over $\lambda\in[0,T_{\rm orbit}]$ (or $3T_{\rm orbit}$ for the irrational case). Figure~\ref{fig:rosettes} shows the resulting closed rosette orbits, projected onto the orbital plane via $x=r\cos\phi$, $y=r\sin\phi$, for the six rational cases of Table~\ref{tab:orbits}; Fig.~\ref{fig:irrational} shows the corresponding irrational-$q$ orbit. As expected from the zoom-whirl classification, orbits with larger $z$ trace an increasingly petalled rosette with $z$ well-separated ``zoom'' excursions out to apastron, connected by brief  ``whirl'' loops near periastron; the irrational orbit densely fills the annulus bounded by the periastron and apastron circles without ever closing, tracing out the same underlying zoom-whirl skeleton as its rational neighbors.

\subsection{Dependence on the magnetic coupling}
\label{sec:beta-dependence}

Because $\beta$ enters both the effective potential Eq.~\eqref{eq:Veff} and, through $L_{\rm ISCO}(\beta)$, $L_{\rm MBO}(\beta)$, the very construction of the fixed angular momentum used to seed the periodic orbits, it is useful to repeat the construction of Table~\ref{tab:orbits} at several values of $\beta$, holding the spin fixed at $a=0.3$ and re-fixing $L=\tfrac12\big(L_{\rm ISCO}(\beta)+L_{\rm MBO}(\beta)\big)$ at each $\beta$. Table~\ref{tab:beta-orbits} lists the resulting orbital energy $E$ for the same six $(z,w,v)$ configurations as before, at $\beta=0,0.05,0.10,0.15,0.20$; the $\beta=0.10$ row reproduces Table~\ref{tab:orbits} to within numerical precision. As $\beta$ increases, $L$ decreases (following $L_{\rm ISCO}$ and $L_{\rm MBO}$, cf.\ Table~\ref{tab:isco-mbo}) and the orbital energy of every $(z,w,v)$ configuration decreases essentially in lock-step with $E_{\rm MBO}=1-\beta$, while the apastron radius $r_0$ of each orbit increases, exactly as for the individual MBO and ISCO in Sec.~\ref{sec:isco-mbo}.

\begin{table}[ht]
\centering
\caption{Orbital energy $E$ of the same six $(z,w,v)$ periodic-orbit configurations as Table~\ref{tab:orbits}, recomputed at $L=\tfrac12(L_{\rm ISCO}+L_{\rm MBO})$ for several values of $\beta$, at fixed spin $a=0.3$.}
\label{tab:beta-orbits}
\begin{tabular}{@{}ccccccc@{}}
\toprule
$\beta$ & $L/M$ & $E_{(1,1,0)}$ & $E_{(2,1,1)}$ & $E_{(3,1,2)}$ & $E_{(4,1,3)}$ & $E_{(6,1,5)}$ \\
\midrule
0.00 & 3.4135 & 0.957113 & 0.961262 & 0.961616 & 0.961728 & 0.961810 \\
0.05 & 3.2014 & 0.911857 & 0.915692 & 0.916021 & 0.916125 & 0.916201 \\
0.10 & 2.9867 & 0.866554 & 0.870067 & 0.870369 & 0.870463 & 0.870536 \\
0.15 & 2.7687 & 0.821194 & 0.824377 & 0.824653 & 0.824740 & 0.824805 \\
0.20 & 2.5464 & 0.775769 & 0.778610 & 0.778858 & 0.778937 & 0.778996 \\
\bottomrule
\end{tabular}
\end{table}

Figure~\ref{fig:rosette-betascan} (left panel) overlays the $(3,1,2)$ periodic orbit computed at each of these five values of $\beta$, all sharing the same topological zoom-whirl skeleton (three petals, one whirl loop). Increasing $\beta$ expands the whole rosette nearly homologously outward; both the whirl-region periastron and the zoom-region apastron grow with $\beta$ while leaving the qualitative $z=3$ petal structure unchanged; the orbital period lengthens correspondingly, from $T_{\rm orbit}=1005.0\,M$ at $\beta=0$ to $1624.7\,M$ at $\beta=0.20$. This is the periodic-orbit-level manifestation of the outward shift of the ISCO and MBO already seen in Fig.~\ref{fig:isco-mbo}, and, together with Table~\ref{tab:beta-orbits}, provides a direct route to reading off the magnetic coupling $\beta$ from the size and energy of an observed periodic orbit at fixed topological class $(z,w,v)$.

\subsection{Dependence on the BH spin}
\label{sec:spin-dependence}

Because the spacetime is rotating, the same construction can equally be repeated at fixed magnetic coupling while varying the spin $a$, which isolates the purely gravitomagnetic (frame-dragging) contribution to the periodic orbit structure from the electromagnetic one studied above. Table~\ref{tab:spin-orbits} repeats Table~\ref{tab:beta-orbits} at the fiducial coupling $\beta=0.10$, for $a=0,0.3,0.6,0.9$, re-fixing $L=\tfrac12(L_{\rm ISCO}(a)+L_{\rm MBO}(a))$ at each spin.

\begin{table}[h]
\centering
\caption{Orbital energy $E$ of the same six $(z,w,v)$ periodic-orbit configurations as Table~\ref{tab:orbits}, recomputed at $L=\tfrac12(L_{\rm ISCO}+L_{\rm MBO})$ for several values of the spin $a$, at fixed coupling $\beta=0.10$.}
\label{tab:spin-orbits}
\begin{tabular}{@{}ccccccc@{}}
\toprule
$a$ & $L/M$ & $E_{(1,1,0)}$ & $E_{(2,1,1)}$ & $E_{(3,1,2)}$ & $E_{(4,1,3)}$ & $E_{(6,1,5)}$ \\
\midrule
0.0 & 3.2561 & 0.873054 & 0.875254 & 0.875426 & 0.875477 & 0.875514 \\
0.3 & 2.9867 & 0.866554 & 0.870067 & 0.870369 & 0.870463 & 0.870536 \\
0.6 & 2.6497 & 0.853661 & 0.860520 & 0.861219 & 0.861461 & 0.861643 \\
0.9 & 2.1221 & 0.813214 & 0.826997 & 0.830375 & 0.831667 & 0.832754 \\
\bottomrule
\end{tabular}
\end{table}

Unlike the effect of $\beta$, increasing the spin at fixed $\beta$ pulls the whole family of periodic orbits inward: $L$ drops by roughly a third from $a=0$ to $a=0.9$ as both $L_{\rm ISCO}$ and $L_{\rm MBO}$ shrink through ordinary frame dragging (Sec.~\ref{sec:isco-mbo}), and the orbital energies of every $(z,w,v)$ configuration decrease and spread apart e.g., for $(2,1,1)$ versus $(1,1,0)$ the energy gap grows from $2.2\times10^{-3}$ at $a=0$ to $1.38\times10^{-2}$ at $a=0.9$  reflecting the fact that the effective potential becomes more sharply peaked, rather than uniformly rescaled, as the ISCO and MBO are dragged closer to the horizon. This is shown directly in Fig.~\ref{fig:rosette-betascan} (right panel), which overlays the $(3,1,2)$ orbit at $a=0,0.3,0.6,0.9$: the rosette contracts markedly with increasing spin, its apastron shrinking from $r_0=23.96\,M$ at $a=0$ to $7.91\,M$ at $a=0.9$, while the orbital period shortens from $T_{\rm orbit}=1648.0\,M$ to $348.0\,M$ over the same range more than a factor of four since the same topological class is now traced out much closer to the (also shrinking) horizon. In this sense, spin acts oppositely to, and far more efficiently than, the magnetic coupling: $\beta$ gently inflates periodic orbits while leaving their timescale comparatively unchanged, whereas $a$ compresses them substantially in both size and duration, exactly as expected for a rotating, frame-dragging background rather than a purely electromagnetic modification of it.

\section{Gravitational waveforms from periodic orbits}
\label{sec:gw}

\subsection{Numerical-kludge quadrupole waveform}

\begin{figure}[t]
\centering
\includegraphics[width=0.98\linewidth]{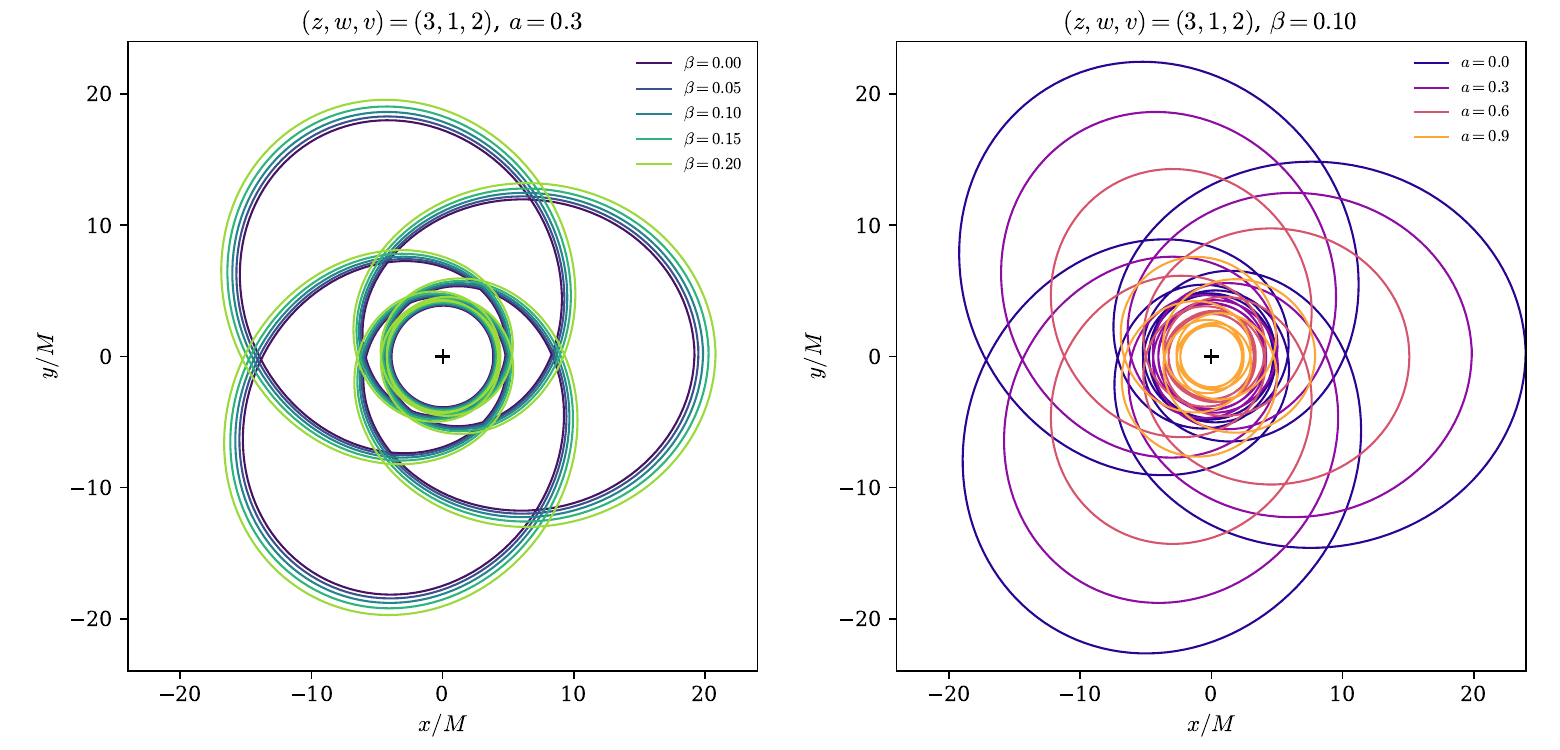}
\caption{The $(z,w,v)=(3,1,2)$ periodic orbit, recomputed at $L=\tfrac12(L_{\rm ISCO}+L_{\rm MBO})$ for $\beta=0,0.05,0.10,0.15,0.20$ at fixed $a=0.3$ (left, Table~\ref{tab:beta-orbits}) and for $a=0,0.3,0.6,0.9$ at fixed $\beta=0.10$ (right, Table~\ref{tab:spin-orbits}). Increasing $\beta$ expands the rosette outward; increasing $a$ instead contracts it inward through frame dragging, in both cases preserving its zoom-whirl topology.}
\label{fig:rosette-betascan}
\end{figure}

We now compute the gravitational radiation produced by the periodic orbits of Sec.~\ref{sec:periodic} using the numerical-kludge, restricted (leading-order) mass-quadrupole approximation~\cite{Peters1963,Thorne1980,Xamidov:2026kqs}, in which the smallness of the mass ratio justifies treating the orbit as adiabatically geodesic-like over the short time intervals (a handful of orbital periods) considered here. We consider an EMRI system consisting of a magnetized neutron-star secondary of mass $m_p$ orbiting a supermassive magnetized Kerr BH of mass $M_\bullet$ at luminosity distance $D$; for concreteness we adopt a secondary mass in the astrophysically realistic $2$-$4\,\Msun$ neutron-star range, a primary of Sgr~A$^\ast$-like mass, and a fiducial luminosity distance representative of a nearby extragalactic EMRI host rather than the Galactic Centre itself,
\begin{equation}
M_\bullet = 4\times10^{6}\,\Msun, \qquad m_p = 2\,\Msun, \qquad D = 200\ {\rm Mpc},
\end{equation}
with symmetric mass ratio $\eta=M_\bullet m_p/(M_\bullet+m_p)^2=4.9999\times10^{-7}$, and observation angles (inclination and viewing azimuth) $\iota=\zeta=\pi/4$. Given the orbit table $\{\lambda_i,\phi_i,r_i\}$ from Sec.~\ref{sec:periodic}, physical time is $t_i=\mathcal{T}\lambda_i$ with $\mathcal T \equiv GM_\bullet/c^3=19.7$~s, and the leading-order quadrupole strain polarizations are~\cite{Alloqulov:2025ucf,Lu:2025xlp}
\begin{equation}
h_+(t_i) = \big(1+\cos^2\iota\big)\cos\!\big(2\phi_i+2\zeta\big)\,\frac{\mathcal A_+}{r_i}\, , \qquad
h_\times(t_i) = \cos\iota\,\sin\!\big(2\phi_i+2\zeta\big)\,\frac{\mathcal A_\times}{r_i}\, ,
\label{eq:hplus-hcross}
\end{equation}
with amplitude scales
\begin{equation}
\mathcal A_+ = -\frac{2\eta}{D}\,\frac{GM_\bullet}{c^2} = -9.536\times10^{-22}, \qquad
\mathcal A_\times = -\frac{4\eta}{D}\,\frac{GM_\bullet}{c^2} = -1.907\times10^{-21},
\end{equation}
where $r_i$ is measured in units of $M_\bullet$. Equation~\eqref{eq:hplus-hcross} is the standard generalized Peters-Mathews quadrupole formula for an eccentric two-body orbit, evaluated using the instantaneous orbital separation and phase from the numerically integrated periodic orbit rather than by finite-differencing the Cartesian quadrupole moment, and is equivalent to projecting the symmetric trace-free mass-quadrupole radiation formula $h_{ij}=(2/D)\ddot I_{ij}$ onto the standard $+/\times$ polarization basis for a source viewed at inclination $\iota$~\cite{Thorne1980}.

\begin{figure}[t]
\centering
\includegraphics[width=0.85\linewidth]{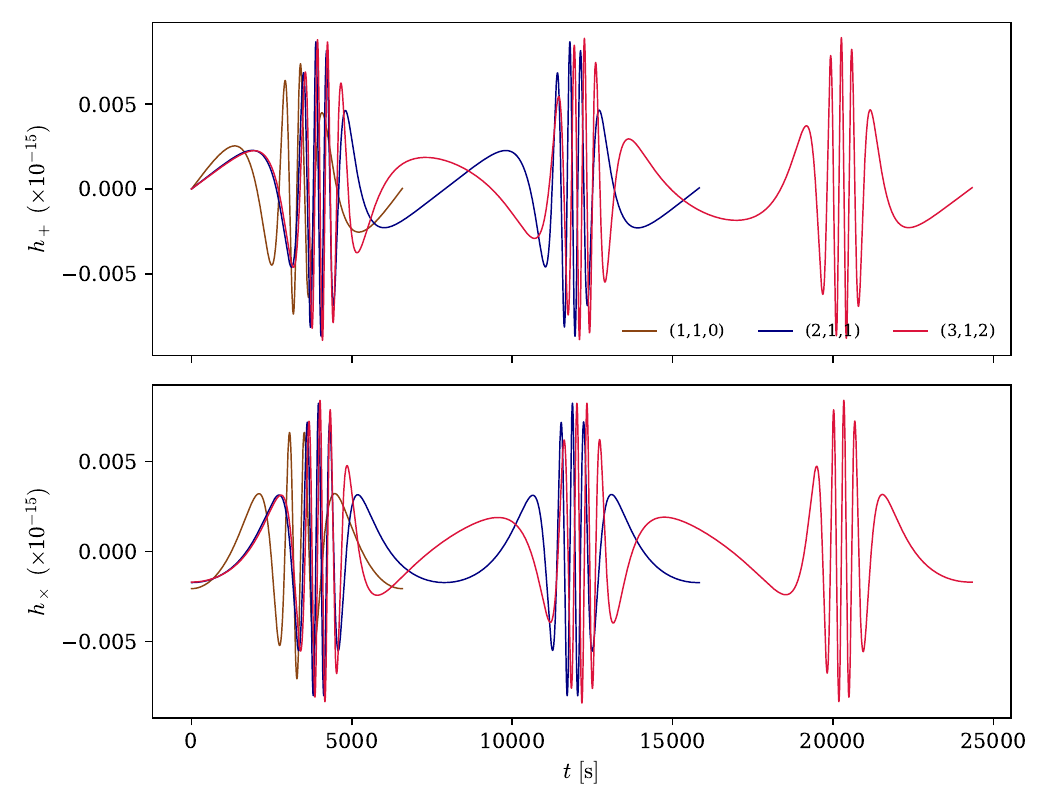}
\caption{Time-domain gravitational waveforms $h_+(t)$ (top) and $h_\times(t)$ (bottom) radiated by the $(1,1,0)$, $(2,1,1)$ and $(3,1,2)$ periodic orbits of Table~\ref{tab:orbits}, for an EMRI with $M_\bullet=4\times10^6\,\Msun$, $m_p=2\,\Msun$, $D=200$~Mpc, $\iota=\zeta=\pi/4$.}
\label{fig:waveform}
\end{figure}

Figure~\ref{fig:waveform} shows the resulting $h_+(t)$ and $h_\times(t)$ for the $(1,1,0)$, $(2,1,1)$ and $(3,1,2)$ periodic orbits. As anticipated from the orbital morphology of Fig.~\ref{fig:rosettes}, each waveform consists of long, quiescent, low-amplitude ``zoom'' segments -- radiated while the particle is far from the BH, near apastron -- punctuated by short, high-frequency, high-amplitude ``whirl'' bursts radiated while the particle loops near periastron just outside the unstable circular orbit. Orbits with larger $z$ produce more numerous, more widely spaced whirl bursts within a single orbital period, directly encoding the zoom-whirl integers $(z,w,v)$ in the time-domain waveform morphology, and, more generally, for periodic orbits in Kerr and modified gravity backgrounds~\cite{LevinPerezGiz2008b,Lin2020}.

\begin{figure}[t]
\centering
\includegraphics[width=0.98\linewidth]{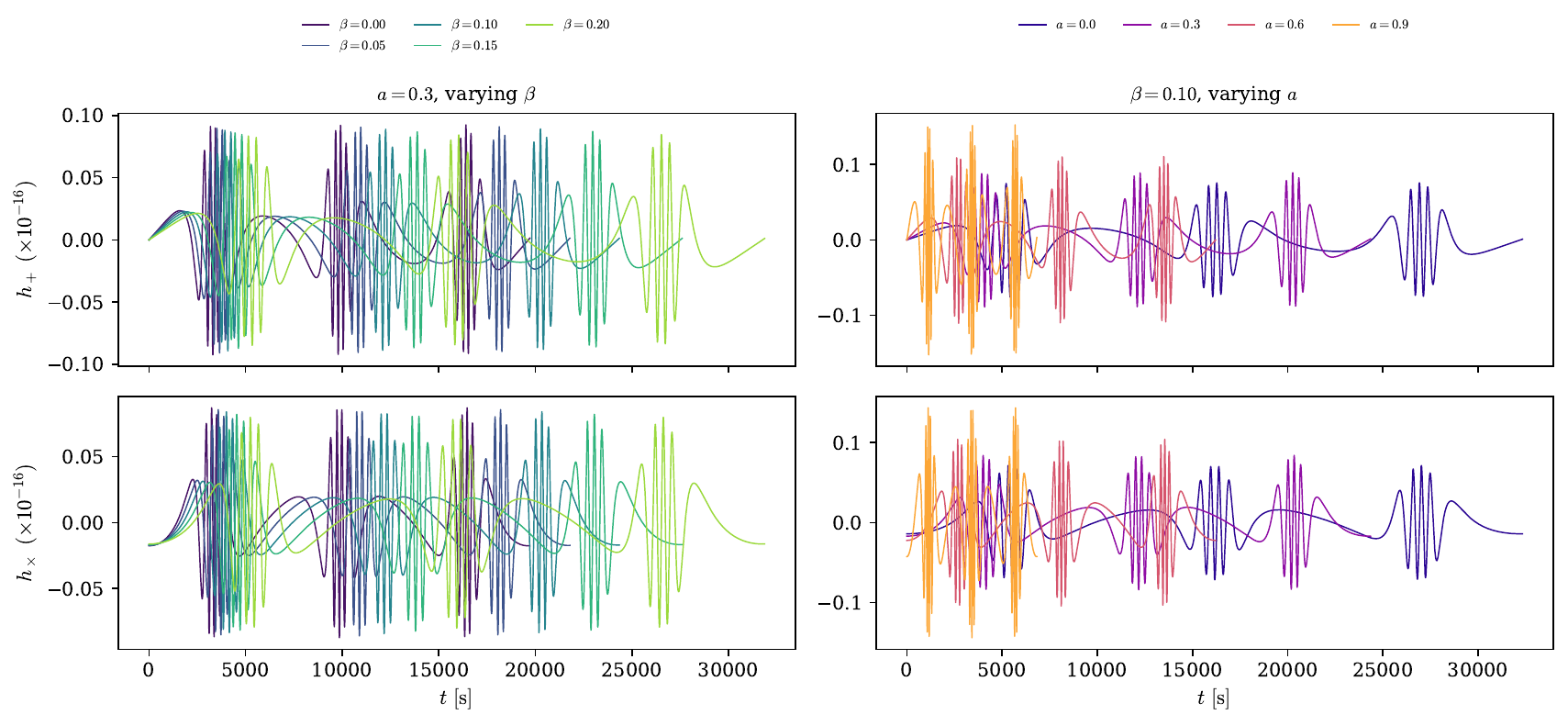}
\caption{Time-domain gravitational waveforms $h_+(t)$ (top row) and $h_\times(t)$ (bottom row) radiated by the single $(z,w,v)=(3,1,2)$ periodic orbit of Fig.~\ref{fig:rosette-betascan}, recomputed at $\beta=0,0.05,0.10,0.15,0.20$ at fixed $a=0.3$ (left column, Table~\ref{tab:beta-orbits}) and at $a=0,0.3,0.6,0.9$ at fixed $\beta=0.10$ (right column, Table~\ref{tab:spin-orbits}), for the same EMRI parameters as Fig.~\ref{fig:waveform}. Increasing $\beta$ stretches the waveform in time at nearly fixed amplitude; increasing $a$ instead compresses it in time while raising the peak strain.}
\label{fig:waveform-betascan}
\end{figure}

To isolate the imprint of the magnetic coupling on the waveform itself, Fig.~\ref{fig:waveform-betascan} (left column) recomputes $h_+(t)$ and $h_\times(t)$ for the single topological class $(z,w,v)=(3,1,2)$ at the five values of $\beta$ of Table~\ref{tab:beta-orbits} (i.e., for the orbits already shown overlaid in Fig.~\ref{fig:rosette-betascan}), fixing all EMRI scaling parameters. Increasing $\beta$ lengthens the zoom-phase duration and delays the onset of each whirl burst -- a direct consequence of the larger apastron and longer orbital period at larger $\beta$ seen in Table~\ref{tab:beta-orbits} -- while leaving the peak strain amplitude and burst count within a cycle essentially unchanged, since these are set by the shared topological class $(z,w,v)=(3,1,2)$ rather than by $\beta$ itself. In this sense $\beta$ acts primarily as a timing parameter (stretching the zoom-whirl waveform along the time axis) rather than an amplitude parameter, which is consistent with its effect on the underlying orbit (Fig.~\ref{fig:rosette-betascan}) being an almost homologous expansion at fixed topology.

The right column of Fig.~\ref{fig:waveform-betascan} repeats this exercise varying the spin instead, at fixed $\beta=0.10$, for the $(3,1,2)$ orbits of Table~\ref{tab:spin-orbits} and Fig.~\ref{fig:rosette-betascan} (right panel). The spin dependence is the mirror image of, and considerably stronger than, the $\beta$ dependence: increasing $a$ compresses the waveform in time the total signal duration shrinks from $t_{\max}=3.23\times10^4$~s at $a=0$ to $6.8\times10^3$~s at $a=0.9$, a factor of $\sim4.7$, tracking the shortening orbital period of Table~\ref{tab:spin-orbits} while simultaneously raising the peak strain amplitude, from $|h_+|_{\max}=7.58\times10^{-18}$ at $a=0$ to $1.53\times10^{-17}$ at $a=0.9$, because the same topological class is traced at a smaller physical radius $r_i$ and the quadrupole amplitude in Eq.~\eqref{eq:hplus-hcross} scales as $1/r_i$. Thus, while $\beta$ leaves an almost purely timing-like imprint on the waveform, the spin $a$ imprints both a timing and an amplitude signature -- a higher, more tightly packed sequence of whirl bursts -- which is the expected signature of a genuinely rotating, frame-dragging background rather than of an added electromagnetic coupling, and in principle allows $a$ and $\beta$ to be disentangled from the morphology of an observed EMRI waveform alone.

\subsection{Frequency domain: characteristic strain and detectability}

The frequency-domain content of the waveform is obtained via a discrete Fourier transform of the (Hann-windowed) time series~\cite{Zahra:2025tdo,Alloqulov:2025tdy,Jumaniyozov:2026bee,Shokirov:2026wez},
\begin{equation}
\tilde h_{+,\times}(f) = \int h_{+,\times}(t)\,e^{-2\pi i f t}\,\dd t \; \xrightarrow{\ {\rm DFT}\ }\; \tilde h_{+,\times}(f_k),
\end{equation}
from which we form the characteristic strain, following the standard convention used e.g.\ in~\cite{Alloqulov2026,Robson2019},
\begin{equation}
h_c(f) = 2f\sqrt{\big|\tilde h_+(f)\big|^2+\big|\tilde h_\times(f)\big|^2}\, .
\label{eq:hc}
\end{equation}

For the fiducial EMRI parameters above, the closed periodic orbits of Table~\ref{tab:orbits} have fundamental orbital periods ranging from $T_{\rm leaf}\mathcal T\simeq6.57\times10^3$~s ($z=1$) to $4.96\times10^4$~s ($z=6$), placing their characteristic gravitational-wave frequencies in the milli-to-deci-Hz band accessible to space-based detectors: we find spectral content spanning approximately $[1.5\times10^{-4},\,0.61]$~Hz for $z=1$ down to $[2.0\times10^{-5},\,0.08]$~Hz for $z=6$ (Table~\ref{tab:gw-summary}), with the low-frequency edge set by the fundamental orbital frequency $1/(\,T_{\rm orbit}\mathcal T)$ and the high-frequency content generated by the sharp whirl bursts.

\begin{table}[t]
\centering
\caption{Physical orbital period, gravitational-wave frequency range, and peak characteristic strain for the six rational periodic orbits, for the EMRI parameters of Fig.~\ref{fig:waveform}.}
\label{tab:gw-summary}
\begin{tabular}{@{}cccc@{}}
\toprule
$(z,w,v)$ & $T_{\rm orbit}\mathcal T$ [s] & $f$ range [Hz] & $h_c^{\max}$ \\
\midrule
$(1,1,0)$ & $6.57\times10^{3}$ & $[1.5\times10^{-4},\ 6.1\times10^{-1}]$ & $1.2\times10^{-21}$ \\
$(2,1,1)$ & $1.58\times10^{4}$ & $[6.3\times10^{-5},\ 2.5\times10^{-1}]$ & $1.9\times10^{-21}$ \\
$(3,1,2)$ & $2.43\times10^{4}$ & $[4.1\times10^{-5},\ 1.6\times10^{-1}]$ & $3.1\times10^{-21}$ \\
$(4,1,3)$ & $3.28\times10^{4}$ & $[3.1\times10^{-5},\ 1.2\times10^{-1}]$ & $4.4\times10^{-21}$ \\
$(5,1,4)$ & $4.12\times10^{4}$ & $[2.4\times10^{-5},\ 9.7\times10^{-2}]$ & $5.7\times10^{-21}$ \\
$(6,1,5)$ & $4.96\times10^{4}$ & $[2.0\times10^{-5},\ 8.1\times10^{-2}]$ & $6.9\times10^{-21}$ \\
\bottomrule
\end{tabular}
\end{table}

\begin{figure}[t]
\centering
\includegraphics[width=0.98\linewidth]{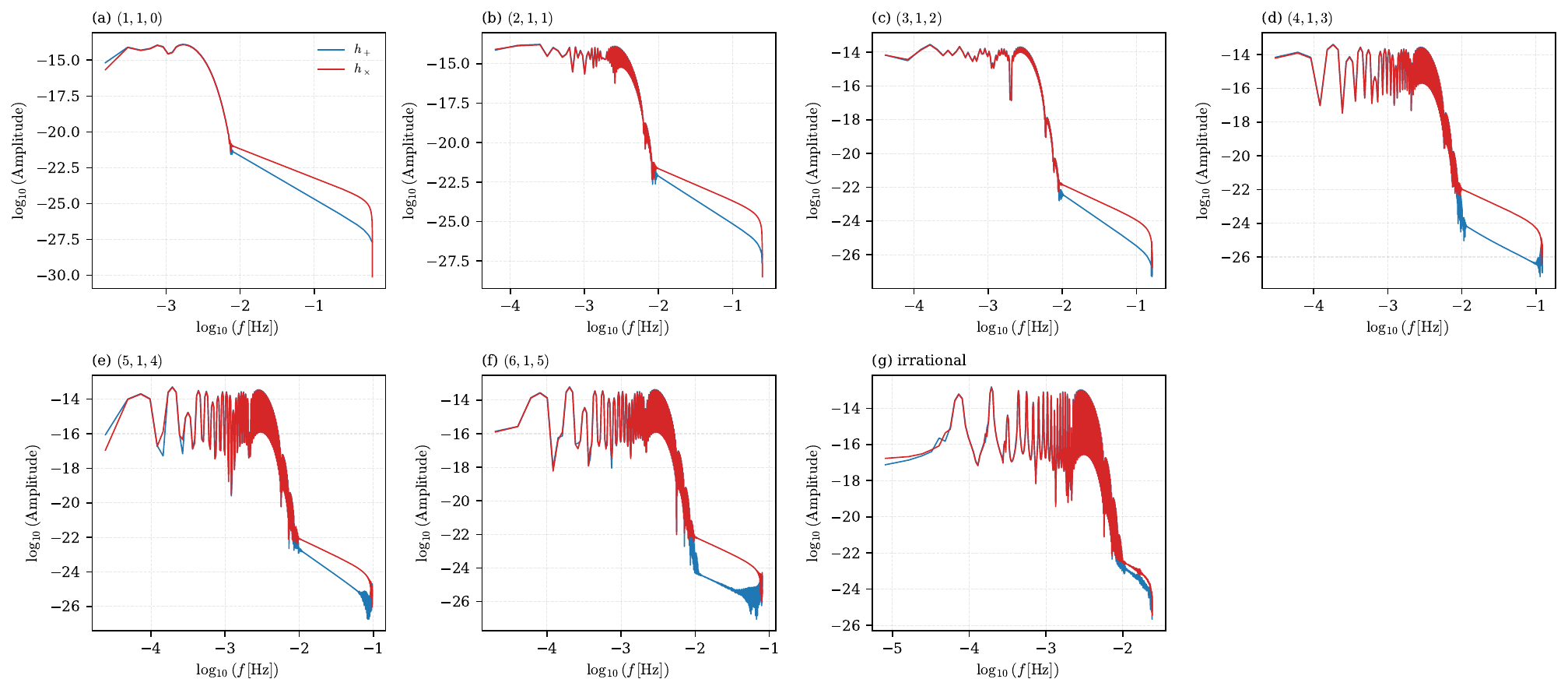}
\caption{Fourier amplitude spectra $|\tilde h_+(f)|$ (blue) and $|\tilde h_\times(f)|$ (red) of the six rational periodic orbits of Table~\ref{tab:orbits}, panels (a)-(f), and of the irrational orbit of Fig.~\ref{fig:irrational}, panel (g), for the EMRI parameters of Fig.~\ref{fig:waveform}.}
\label{fig:gw-fft-panels}
\end{figure}

Figure~\ref{fig:gw-fft-panels} shows the raw Fourier amplitude spectra $|\tilde h_+(f)|$ and $|\tilde h_\times(f)|$ underlying Eq.~\eqref{eq:hc}, individually for each of the six rational periodic orbits of Table~\ref{tab:orbits} and for the irrational orbit of Fig.~\ref{fig:irrational}. Each panel shows the same qualitative shape: a comb of harmonics at the fundamental orbital frequency and its overtones, set by the quiescent zoom phase, followed by a broad excess around $10^{-2}$-$10^{-1}$~Hz sourced by the sharp periastron whirl bursts, before the spectrum falls off steeply once $f$ exceeds the inverse burst duration; increasing $z$ (equivalently, the number of whirl cycles per closed orbit) fills in progressively more overtones between the fundamental and the burst-dominated excess, consistent with the growing number of "petals" seen in the corresponding rosette orbits (Fig.~\ref{fig:rosettes}).

\begin{figure}[t]
\centering
\includegraphics[width=0.68\linewidth]{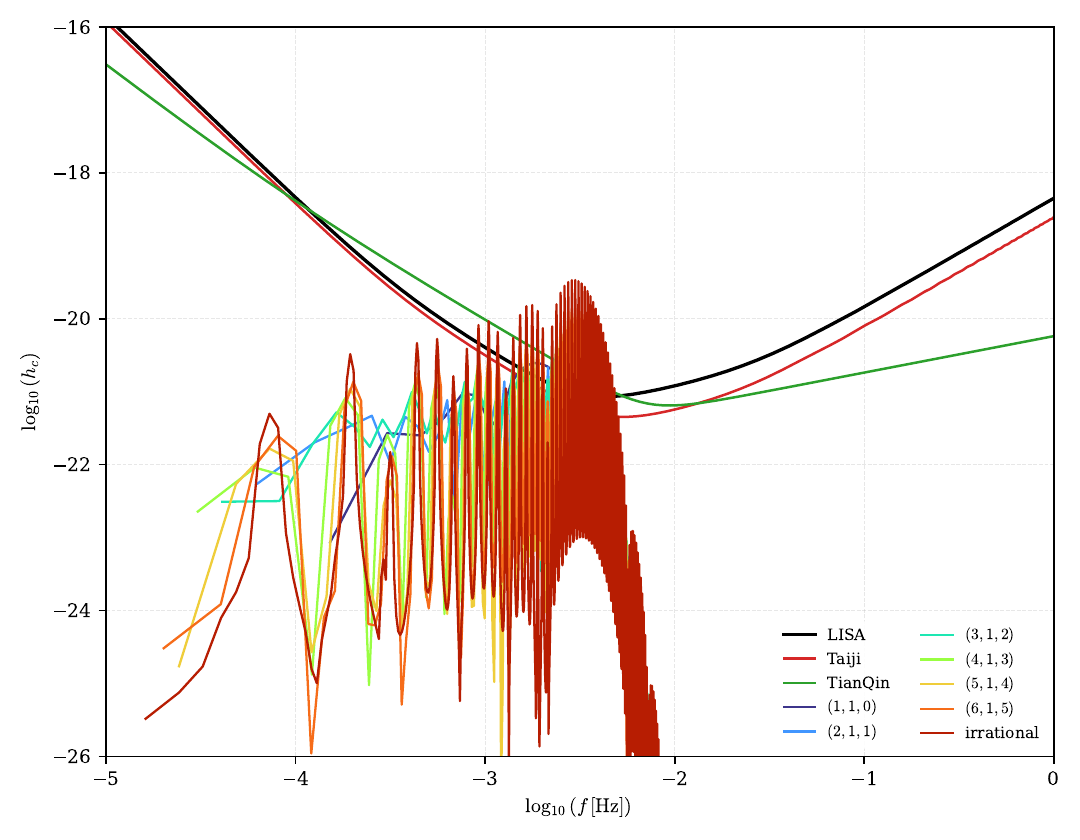}
\caption{Characteristic strain spectra of all six rational periodic orbits $(z,w,v)$ of Table~\ref{tab:orbits} and of the quasi-periodic irrational orbit of Fig.~\ref{fig:irrational}, compared with the analytic instrumental sensitivity curves of LISA, Taiji and TianQin.}
\label{fig:hc-detectors}
\end{figure}

Combining the plus and cross amplitude spectra of Fig.~\ref{fig:gw-fft-panels} via Eq.~\eqref{eq:hc}, Figure~\ref{fig:hc-detectors} compares the characteristic strain spectra of all six rational periodic orbits of Table~\ref{tab:orbits} together with the quasi-periodic irrational orbit of Fig.~\ref{fig:irrational}, against the projected instrumental sensitivity curves of LISA~\cite{Robson2019}, Taiji~\cite{Taiji2020a,Taiji2020b} and TianQin~\cite{TianQin2016}, using standard analytic approximations to their noise power spectral densities. At this fiducial $D=200$~Mpc distance -- representative of a nearby extragalactic EMRI host rather than the Galactic Centre, and roughly four orders of magnitude farther than a Galactic-Centre-like placement of the same source -- the whirl-burst spectral peak no longer sits comfortably above the detector noise floors: it ranges from $h_c^{\max}\simeq1.2\times10^{-21}$ for the $(1,1,0)$ orbit up to $6.9\times10^{-21}$ for the $(6,1,5)$ orbit (Table~\ref{tab:gw-summary}), which is, at the corresponding peak frequency ($\sim2$-$3$~mHz), a factor of $\sim0.8$-$7$ times the LISA noise curve, $\sim1$-$11$ times the Taiji curve, and $\sim0.3$-$4$ times the (locally less sensitive) TianQin curve -- an order-unity peak-to-floor ratio rather than the several-orders-of-magnitude margin obtained for the same construction at a Galactic-Centre-like distance. A real detection at $D=200$~Mpc would therefore require coherently accumulating signal-to-noise over the many burst cycles of the full radiation-reacted inspiral via matched filtering, rather than relying on a single periodic-orbit burst standing clearly above the noise; as throughout this section, the peak-strain-vs-sensitivity-curve comparison shown here is only an illustrative indicator and not a matched-filter SNR calculation. The qualitative trend with $z$ survives the change in distance: more whirl-bursty orbits (higher $z$) radiate more strongly at fixed $(M_\bullet,m_p,D)$, so the $(6,1,5)$ orbit remains the most promising of the six for detectability regardless of how close the source is.

Repeating the same windowed-FFT and characteristic-strain construction for the single $(z,w,v)=(3,1,2)$ orbit of Figs.~\ref{fig:rosette-betascan}-\ref{fig:waveform-betascan}, now scanned over $\beta=0,0.05,0.10,0.15,0.20$ at fixed $a=0.3$ and over $a=0,0.3,0.6,0.9$ at fixed $\beta=0.10$, gives the two panels of Fig.~\ref{fig:hc-betascan}. Increasing $\beta$ at fixed spin leaves the whirl-burst spectral peak at nearly the same characteristic strain while shifting it to slightly lower frequency, tracking the lengthening of the orbital period already seen in Fig.~\ref{fig:waveform-betascan}: the magnetic coupling redistributes power in frequency without appreciably changing the peak $h_c$. Increasing the spin $a$ instead shifts the peak strongly toward higher frequency by more than an order of magnitude in $f$ from $a=0$ to $a=0.9$ -- while simultaneously raising the peak characteristic strain by a comparable factor, consistent with the frame-dragging-driven contraction and speed-up of the orbit noted above. In both scans the whirl-burst peak stays within an order of magnitude of the LISA/Taiji/TianQin sensitivity curves in their most sensitive band from $\sim1$-$4\times$ the LISA curve and $\sim1$-$10\times$ the Taiji curve across the full $\beta$ and $a$ ranges probed, with the TianQin comparison in particular crossing from below to above unity as $a$ increases -- reflecting the same order-unity peak-to-floor regime found for the fixed-orbit scan of Fig.~\ref{fig:hc-detectors} at this $D=200$~Mpc distance, but the qualitatively different way $\beta$ and $a$ move the spectral peak -- an almost pure frequency shift at fixed amplitude for $\beta$, versus a combined frequency-and-amplitude shift for $a$ -- gives a second, frequency-domain handle (complementing the time-domain morphology discussed above) for disentangling the two parameters from an observed zoom-whirl spectrum.

\section{Conclusions}
\label{sec:conclusions}

\begin{figure}[t]
\centering
\includegraphics[width=0.98\linewidth]{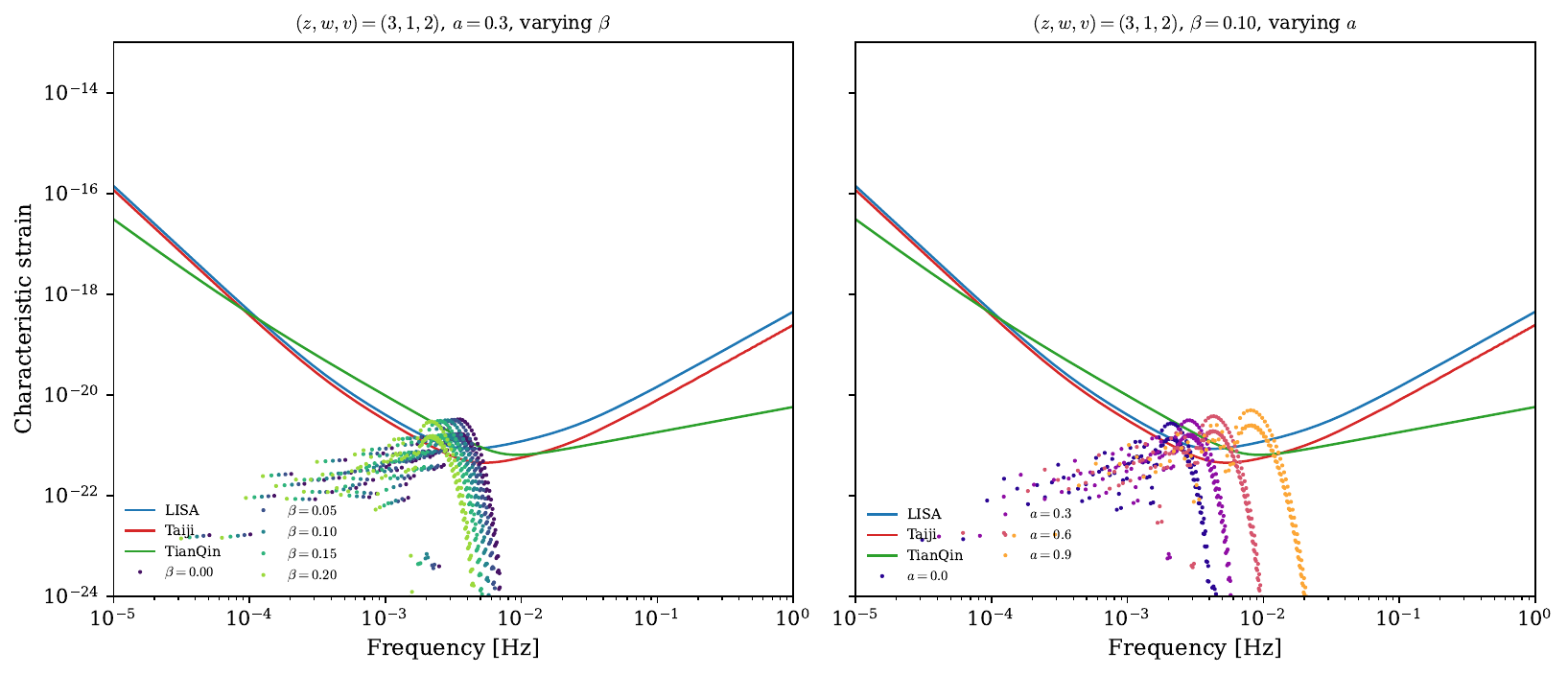}
\caption{Characteristic strain spectrum of the $(z,w,v)=(3,1,2)$ periodic orbit, recomputed at $\beta=0,0.05,0.10,0.15,0.20$ at fixed $a=0.3$ (left) and at $a=0,0.3,0.6,0.9$ at fixed $\beta=0.10$ (right), compared with the analytic instrumental sensitivity curves of LISA, Taiji and TianQin. Increasing $\beta$ shifts the whirl-burst peak to lower frequency at nearly fixed amplitude; increasing $a$ shifts it to higher frequency while also raising the peak strain.}
\label{fig:hc-betascan}
\end{figure}

We have studied periodic orbits and their gravitational wave signatures for a magnetized, chargeless test particle moving in the equatorial plane of a rotating BH immersed in an external magnetic field. Our main findings are:

\begin{itemize}
\item The generalized escape energy of a magnetized particle, $E_{\rm esc}=|1-\beta|$, together with the modified effective potential Eq.~\eqref{eq:Veff}, shifts both the marginally bound orbit and the innermost stable circular orbit outward and lowers their orbital energy and angular momentum as the magnetic coupling $\beta$ increases (Table~\ref{tab:isco-mbo}, Fig.~\ref{fig:isco-mbo})  the opposite trend to that produced by increasing electric charge in the dyonic ModMax BH~\cite{Alloqulov2026}, and consequently a potentially distinguishing observational signature of a magnetic, rather than purely electric, deviation from vacuum Kerr.
\item The allowed region of the orbital angular-momentum-energy plane for bound motion (Fig.~\ref{fig:bound-region}), and the rational rotation number $q(E,L)$ that classifies periodic orbits (Fig.~\ref{fig:qE}), both shift systematically with $\beta$, providing, in principle, an independent handle on the magnetic coupling from the frequency-domain structure of an observed periodic-orbit waveform.
\item Periodic orbits constructed at fixed angular momentum and increasing zoom number $z$ (Table~\ref{tab:orbits}, Fig.~\ref{fig:rosettes}) display the expected zoom-whirl rosette morphology, with the number of "petals" set by $z$; the corresponding gravitational waveforms (Fig.~\ref{fig:waveform}) directly encode this structure as an alternating sequence of quiescent zoom phases and high-frequency whirl bursts, whose frequency-domain content (Fig.~\ref{fig:hc-detectors}, Table~\ref{tab:gw-summary}) spans the milli-to-deci-Hz band targeted by LISA, Taiji and TianQin.
\item Repeating the same $(z,w,v)=(3,1,2)$ orbit construction at $\beta=0,0.05,0.10,0.15,0.20$ (Sec.~\ref{sec:beta-dependence}, Table~\ref{tab:beta-orbits}, Figs.~\ref{fig:rosette-betascan}-\ref{fig:waveform-betascan}) shows that the magnetic coupling acts on a fixed topological class of periodic orbit almost homologously: it expands the rosette outward, lowers its orbital energy in lock-step with $E_{\rm MBO}=1-\beta$, and lengthens the orbital period and the corresponding zoom-phase duration of the waveform, without changing the number of zoom-whirl petals or bursts i.e., $\beta$ acts as a timing/scale parameter at fixed $(z,w,v)$, distinct from the topological information carried by $(z,w,v)$ itself.
\item Because the background is a rotating BH, we additionally repeated this construction at fixed $\beta=0.10$ over $a=0,0.3,0.6,0.9$ (Sec.~\ref{sec:spin-dependence}, Table~\ref{tab:spin-orbits}, Figs.~\ref{fig:rosette-betascan}-\ref{fig:waveform-betascan}) and find that the spin acts essentially oppositely to, and far more strongly than, the magnetic coupling: frame dragging drags $L_{\rm ISCO}$, $L_{\rm MBO}$ and the whole periodic-orbit family inward, contracting the $(3,1,2)$ rosette's apastron by a factor of $\sim3$ and shortening its orbital period by a factor of $\sim4.7$ from $a=0$ to $a=0.9$, while simultaneously raising the peak gravitational-wave strain (by the same factor, since the amplitude scales as the inverse of the physical whirl radius) rather than leaving it essentially unchanged as $\beta$ does. Because $\beta$ imprints an almost pure timing/scale signature at fixed amplitude while $a$ imprints a combined timing-and-amplitude signature, the two parameters are, in principle, separately identifiable from the morphology of a single observed zoom-whirl waveform, without needing to independently measure the BH spin by other means a direct consequence of working with the physically correct rotating (Kerr-based) background rather than a spherically symmetric approximation to it.
\item This time-domain distinction between $\beta$ and $a$ carries over directly to the frequency domain (Fig.~\ref{fig:hc-betascan}): scanning the $(3,1,2)$ orbit's characteristic strain $h_c(f)$ over $\beta$ at fixed $a$ shifts the whirl-burst spectral peak to lower frequency at essentially fixed peak amplitude, while scanning over $a$ at fixed $\beta$ shifts the peak to higher frequency and raises its amplitude by a comparable factor so the two parameters remain separately identifiable from the spectrum alone, without requiring a full time-domain reconstruction of the waveform.
\end{itemize}

Natural extensions of this work include incorporating the resulting radiation-reaction (inspiral) evolution of $(E, L)$ self-consistently across many periodic-orbit cycles, extending the analysis away from the equatorial plane, and confronting the magnetic coupling $\beta$ with the electric-charge and nonlinear-electrodynamics parameters of related magnetized/charged black-hole spacetimes~\cite{Alloqulov2026,Rayimbaev2020} as an observational discriminant for future space-based gravitational-wave missions.

\section*{Acknowledgments}
 J.R. thanks Grant No. F-FA-2021-510 of the Uzbekistan Agency for Innovative Development. Y.C. and J.R. acknowledge the grant of the National Natural Science Foundation of China (NSFC) under Grant No.U2541210.

\section*{Data Availability Statement} This paper is a pure theoretical study, and no associated code/software is involved.

\section*{Code Availability} Code/Software sharing does not apply to this article as no code/software was generated or analyzed during the current study.

\bibliographystyle{apsrev4-1}
\bibliography{references.bib}

\end{document}